\documentclass[12pt]{article}
\usepackage{graphics}
\usepackage{epsfig}
\usepackage{color}
\usepackage{amsmath}
\usepackage{amssymb}
\def\da{\dot{a}}
\def\db{\dot{b}}
\def\G{\Gamma}
\def\D{\Delta}
\def\L{\Lambda}

\def\a{\alpha}
\def\b{\beta}
\def\g{\gamma}
\def\d{\delta}
\def\e{\varepsilon}
\def\m{\mu}
\def\tr{\mbox{tr}}

\def\s{\sigma}
\def\r{\rho}
\def\l{\lambda}
\def\t{\tau}
\def\o{\omega}

\def\vt{\vartheta}
\def\mc{\mathcal}

\def\p{\partial}
\def\la{\langle}
\def\ra{\rangle}
\def\dg{\dagger}
\def\wt{\widetilde}
\numberwithin{equation}{section} 
\parskip=\medskipamount

\def\Zop{\bbbz}

\def\Nop{\bbbn}
\def\bbbz {{\sf Z\!\!Z}}

\def\bbbn {{\rm I\!N}}
\def\RR{R-R }

\begin{document}

\thispagestyle{empty}
\addtocounter{page}{-1}
\def\thefootnote{\fnsymbol{footnote}}
\begin{flushright}
  hep-th/0210246 \\
  AEI-2002-086 \\
  SPIN-2002/33 \\
  ITP-2002/54
\end{flushright}

\vskip 0.5cm

\begin{center}\LARGE
{\bf PP-Wave Light-Cone Superstring Field Theory}
\end{center}

\vskip 1.0cm

\begin{center}
{\large A. Pankiewicz\footnote{E-mail address: {\tt apankie@aei.mpg.de}} and
B. Stefa\'nski, jr.\footnote{E-mail address: {\tt stefansk@phys.uu.nl}}}

\vskip 0.5cm

{\it $^*$ Max-Planck-Institut f\"ur Gravitationsphysik, Albert-Einstein
Institut \\ Am M\"uhlenberg 1, D-14476 Golm, Germany}

\vskip 0.5cm

{\it $^\dagger$ Spinoza Institute, University of Utrecht \\
Postbus 80.195, 3508 TD Utrecht, The Netherlands}
\end{center}

\vskip 1.0cm

\begin{center}
October 2002
\end{center}

\vskip 1.0cm

\begin{abstract}
\noindent 
We construct the cubic interaction vertex and dynamically
generated supercharges in light-cone superstring field theory in the pp-wave background. We show that these satisfy
the pp-wave superalgebra at first order in string coupling. 
The cubic interaction vertex and dynamical supercharges presented here differ from the expressions previously given 
in the literature.
Using this vertex we compute various string theory
three-point functions and comment on their relation to gauge theory in the BMN limit.
\end{abstract}

\vfill

\setcounter{footnote}{0}
\def\thefootnote{\arabic{footnote}}
\newpage

\renewcommand{\theequation}{\thesection.\arabic{equation}}
\section{Introduction and Summary}
\setcounter{equation}{0}

Following earlier work on maximally supersymmetric solutions of 
eleven-dimensional supergravity~\cite{KG,Hull,jffp}, a new maximally supersymmetric
solution of Type IIB supergravity~\cite{bfhp} was found. This paved the way for
quantisation in light-cone gauge of superstring theory in a
constant \RR flux~\cite{m}.
The spectrum consists of a
unique massless groundstate on which a tower of massive states is
constructed~\cite{m,mt} using creation operators whose masses are of order
$\omega_n/(\alpha'p^+)$ where
\begin{equation}
\omega_n\equiv\sqrt{n^2+(\alpha^\prime\mu p^+)^2}\,,\qquad n\in\Nop\,,
\end{equation}
$p^+$ is momentum along the $x^-$ light-cone direction and $\mu$ is the \RR flux. 
This new string background merits further investigation. 

\begin{figure}[htb]
\begin{center}
\vspace{1cm}
\includegraphics{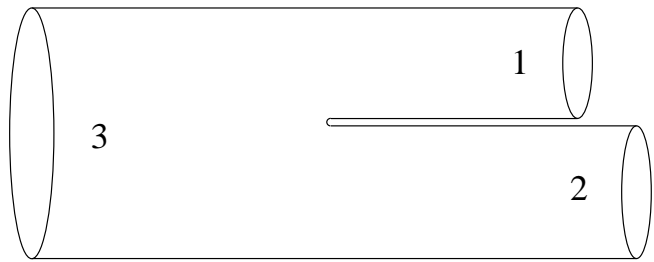}
\end{center}
\caption{The worldsheet of the three string interaction vertex.}
\label{fig1}
\end{figure}

\noindent Since the string spectrum in the pp-wave background is now known, 
string interactions are the next step in the study of the pp-wave background. In this background ten-dimensional 
Lorentz invariance is broken by the 
non-zero \RR flux, and hence it is no longer possible to
set $p^+$ to zero in general scattering amplitudes. This obstruction significantly hinders the
vertex operator approach to string interactions. There is only one other known way of studying string interactions in
light-cone gauge pioneered by Mandelstam~\cite{Mand} for the bosonic string. In this approach
an interaction vertex for the scattering of three strings can be constructed by requiring continuity of string fields
on the worldsheet depicted in Figure~\ref{fig1}. This continuity is enforced by a delta functional
on string fields
\begin{equation}
\Delta(X^1+X^2-X^3)\,.
\end{equation}
For computational purposes it is essential to express the delta functional in Fourier modes~\cite{CG}.
The interaction vertex may then be written as an exponential of creation operators which enforce
the delta functional conditions mode by mode.

\noindent The functional approach~\cite{Mand,KK} can be extended to the superstring~\cite{Mandsusy,gs,gsb}. 
The cubic interaction vertex is a first order in string coupling $g_s$ 
correction to the free Hamiltonian. In order to
satisfy the superalgebra with this new Hamiltonian, dynamical supercharges also receive corrections. This
essential difference from the bosonic string modifies the form of the vertex~\cite{gs,gsb}. In particular~\cite{gsb}, 
in the
oscillator basis, the superstring interaction vertex as well as the dynamical supercharges
not only have a part exponential in creation operators, but
also have a prefactor which is polynomial in creation operators.
The exact form of the prefactors is determined
by two requirements. Firstly, they should not destroy worldsheet continuity enforced by the
exponential part of the vertex.
Secondly, the superalgebra should be satisfied order by order in string coupling.
In~\cite{gsb} an oscillator expression for the flat spacetime superstring vertex and dynamical supercharges was 
constructed and shown to satisfy all said consistency conditions.

\noindent In this paper we apply the formalism of~\cite{CG,gs,gsb} to the pp-wave background in order
to construct an interaction vertex and dynamical supercharges in the oscillator basis. We prove explicitly that our
expressions satisfy all the aforementioned consistency conditions. We use this vertex to compute various
three-point functions in string theory.

\noindent The pp-wave interaction vertex has also been recently investigated in~\cite{sv,sv2}. Our results differ
from the expressions presented in~\cite{sv,sv2} in several ways. Firstly, our bosonic prefactor contains 
a piece, not present in~\cite{sv,sv2}, proportional to $\mu\delta^{IJ}$ ({\it cf}. equation~\eqref{h}). 
This piece is crucial in proving (see section~\ref{sec4}) that
the dynamically generated supercharges and interaction Hamiltonian satisfy the pp-wave superalgebra at first
order in string coupling. Further, without it, three-point functions of states containing only fermionic
creation operators would vanish (see section~\ref{sec52}). Secondly, we also present the oscillator-basis 
expression for the fermionic prefactor; in particular, the $\m$-dependent normalisation ({\it cf}. equation~\eqref{y}) 
is essential in 
proving that the dynamical superalgebra is satisfied. The exponential part of our vertex also contains a piece linear 
in fermionic zero-modes~\cite{ap} ({\it cf}. equation~\eqref{fv}) and our expression for the
coefficients of the negatively moded bosonic creation operators in the prefactor ({\it cf}. equation~\eqref{negf})
differs by a factor of $i$~\cite{ap} from the original one in~\cite{sv2}.\footnote{Recently a new version of~\cite{sv2} 
has appeared correcting the factor of $i$ discrepancy.} 
Since we prove that our expressions for the dynamical supercharges and Hamiltonian satisfy the superalgebra, we
believe that the vertex constructed here is the correct one. 

\noindent The pp-wave background can be obtained as a Penrose limit~\cite{penrose} of
$AdS_5\times S^5$~\cite{bfhp}. Through the AdS/CFT correspondence~\cite{malda,GKP,Witten}, this has 
led to a conjectured duality between string theory in the pp-wave background and
a double scaling limit of SU($N$), ${\mc N}=4$ Super Yang-Mills theory~\cite{bmn}.
Explicitly one considers a sector of (BMN) operators with $U(1)$ R-charge $J$ and
conformal dimension $\D$ such that the difference $\D-J$ remains fixed in
the limit $J$, $N\to\infty$, $J^2/N$ and $g_{\mbox{\scriptsize YM}}$ fixed~\cite{bmn}. 
The BMN operators are field theory duals of perturbative string states in the pp-wave background. 
The SYM theory is believed to be expandable in a double series in the 
effective genus counting parameter $g_2^2$ and 
effective coupling $\lambda'$~\cite{jan,freedman}
\begin{equation}
g_2^2=\frac{J^4}{N^2}\,,\qquad\lambda'=\frac{g^2_{\mbox{\scriptsize YM}}N}{J^2}\,.
\end{equation}
Moreover, the finite quantity $\D-J$ is a function of $\lambda'$~\cite{bmn,gross,sz} 
and $g_2^2$~\cite{jan, freedman}. 
The expansion parameters $g_2$ and $\lambda'$ are related to string theory parameters by~\cite{bmn}
\begin{equation}
\frac{1}{(\m\a'p^+)^2}=\lambda'\,,\qquad 4\pi g_{\text{s}}(\m\a'p^+)^2=g_2\,.
\end{equation}
One of the most appealing aspects of this duality is that both string theory and the BMN sector of gauge theory are 
simultaneously perturbatively accessible.

\noindent
While some further investigation seems necessary to clarify the exact form of this correspondence, at the
planar (interacting) field theory two-point function/free string level, there is increasing 
evidence~\cite{bmn,gross,sz} 
that the BMN operators do correspond to the free string theory. The extension of the duality
beyond this limit, by considering string interactions and the non-planar sector of (interacting) gauge theory 
respectively, 
needs further clarification. Indeed, the exact notion of the Penrose limit in a CFT needs to be better 
understood~\cite{AS}.
By focusing on a certain subclass of the BMN operators, several proposals for such an extension of the BMN
correspondence have been made and were studied in~\cite{jan,freedman,verlinde,zhou,jan2,freedman2,vv,janik,vsv,gomis}. 
In this paper we use our interaction vertex to compute
three-point functions for various string states. In particular, for the class of amplitudes studied so far in the 
literature the term in the cubic vertex proportional to $\m\d^{IJ}$ does not contribute 
and we recover the results of~\cite{sv2,vsv,gomis}. We show that so-called impurity preserving 
three-point functions of states dual to a class of protected operators vanish in string theory, as would be expected 
by the extension of the duality recently proposed in~\cite{vsv,gomis}. 
Further, we compute three-point functions for string states having only fermionic creation operators. Here 
only the $\m\d^{IJ}$ term contributes and these amplitudes may provide a check of any extension of the 
string-bit dynamical supercharges to terms bilinear in fermions, which are not known at present~\cite{vv}.
We hope our results can shed some light on the BMN correspondence and that the three-string vertex can be used to 
investigate the duality further.
Finally if the BMN duality is to be believed, D-branes in the pp-wave background~\cite{skenderis,bgg}
should play an important role in gauge theory.

\noindent This paper is organized as follows. In section~\ref{sec2} we briefly review
the free string in the pp-wave background and set our notation. In sections~\ref{sec3} and~\ref{sec4} we
construct the superstring vertex in the pp-wave background. Section~\ref{sec3} focuses on the exponential part of
the vertex. Most of the results of this section have appeared in~\cite{sv,sv2,ap} and are included
here for completeness.
In section~\ref{sec4} we construct the prefactors and show that the superalgebra is satisfied with these modified
generators. Using the vertex, we compute some three-point functions and compare them to field theory in the BMN limit
in section~\ref{sec5}. 
Several appendices, containing computational details are also included.
\section{Review of free string theory on the pp-wave}\label{sec2}
The free string in the maximally supersymmetric pp-wave background in
light-cone gauge is described by $x^I_r(\s_r)$ and $\vt^a_r(\s_r)$ in position space or
by $p^I_r(\s_r)$ and $\l^a_r(\s_r)$ in momentum space, where $I$ is a transverse
SO(8) vector index, $a$ is a SO(8) spinor index.\footnote{We will often suppress these
indices in what follows.} The index $r=1,2,3$ denotes the $r$th
string (see Figure~\ref{fig1}). The bosonic part of the light-cone action in the pp-wave
background is~\cite{m}
\begin{equation}
S_{\mbox{\scriptsize bos.}(r)}=\frac{e(\a_r)}{4\pi\a'}\int\,d\t\int_0^{2\pi|\a_r|}
\,d\s_r\bigl[\dot{x}_r^2-x^{\prime\,2}_r-\m^2x_r^2\bigr]\,,
\end{equation}
where
\begin{equation}
\dot{x}_r\equiv\p_{\t}x_r\,,\qquad x'_r\equiv\p_{\s_r}x_r\,,\qquad
\a_r\equiv\a'p^+_r\,,\qquad e(\a_r)\equiv\frac{\a_r}{|\alpha_r|}\,.
\end{equation}
In a collision process $p^+_r$ will be positive for an incoming string and
negative for an outgoing one.
The mode expansions of the fields $x_r^I(\s_r,\t)$ and $p_r^I(\s_r,\t)$ at $\t=0$
are
\begin{equation}
\begin{split}
x_r^I(\s_r)& = x_{0(r)}^I+\sqrt{2}\sum_{n=1}^{\infty}
\bigl(x_{n(r)}^I\cos\frac{n\s_r}{|\a_r|}+x_{-n(r)}^I\sin\frac{n\s_r}{|\a_r|}\bigr)\,,\\
p_r^I(\s_r) & =\frac{1}{2\pi|\a_r|}\bigl[p_{0(r)}^I+\sqrt{2}\sum_{n=1}^{\infty}
\bigl(p_{n(r)}^I\cos\frac{n\s_r}{|\a_r|}+p_{-n(r)}^I\sin\frac{n\s_r}{|\a_r|}\bigr)\bigr]\,.
\end{split}
\end{equation}
The Fourier modes can be re-expressed in terms of creation and
annihilation operators as
\begin{equation}\label{xp}
x_{n(r)}^I=i\sqrt{\frac{\a'}{2\o_{n(r)}}}\bigl(a_{n(r)}^I-a_{n(r)}^{I\,\dg}\bigr)\,,\qquad
p_{n(r)}^I=\sqrt{\frac{\o_{n(r)}}{2\a'}}\bigl(a_{n(r)}^I+a_{n(r)}^{I\,\dg}\bigr)\,.\qquad
\end{equation}
Canonical quantization of the bosonic coordinates
\begin{equation}
[x_r^I(\s_r),p_s^J(\s_s)]=i\d^{IJ}\d_{rs}\d(\s_r-\s_s)
\end{equation}
yields the usual commutation relations
\begin{equation}
[a_{n(r)}^I,a_{m(s)}^{J\,\dg}]=\d^{IJ}\d_{nm}\d_{rs}\,.
\end{equation}
The fermionic part of the light-cone action in the pp-wave background is~\cite{m}
\begin{equation}
S_{\mbox{\scriptsize ferm.}(r)}=\frac{1}{8\pi}\int\,d\t\int_0^{2\pi|\a_r|}\,d\s_r
[i(\bar{\vt}_r\dot{\vt}_r+\vt_r\dot{\bar{\vt}}_r)
-\vt_r\vt'_r+\bar{\vt}_r\bar{\vt}'_r-2\m\bar{\vt}_r\Pi\vt_r]\,,
\end{equation}
where $\vt^a_r$ is a complex, positive chirality SO(8) spinor and
\begin{equation}
\Pi_{ab}\equiv(\g^1\g^2\g^3\g^4)_{ab}
\end{equation}
is symmetric, traceless and squares to one.\footnote{Throughout this paper we use
the gamma matrix conventions of~\cite{gsb}. It is convenient to use a representation of gamma matrices in which 
$\Pi_{ab}=\text{diag}({\bf 1}_4\,,-{\bf 1}_4)$.}
The mode expansion of $\vt^a_r$ and its conjugate momentum
\begin{equation}
i\l^a_r\equiv \frac{\delta S_{\mbox{\scriptsize ferm.}(r)}}{\delta \dot{\vt}_{a(r)}}=
i\frac{1}{4\pi}\bar{\vt}^a_r\,,
\end{equation}
at $\t=0$ is
\begin{equation}
\begin{split}
\vt^a_r(\s_r) & =\vt^a_{0(r)}+\sqrt{2}\sum_{n=1}^{\infty}
\bigl(\vt^a_{n(r)}\cos\frac{n\s_r}{|\a_r|}+\vt^a_{-n(r)}\sin\frac{n\s_r}{|\a_r|}\bigr)\,,\\
\l^a_r(\s_r) & =\frac{1}{2\pi|\a_r|}\bigl[\l^a_{0(r)}+\sqrt{2}\sum_{n=1}^{\infty}
\bigl(\l^a_{n(r)}\cos\frac{n\s_r}{|\a_r|}+\l^a_{-n(r)}\sin\frac{n\s_r}{|\a_r|}\bigr)\bigr]\,.
\end{split}
\end{equation}
The Fourier modes satisfy the reality condition
\begin{equation}
\l_{n(r)}^a=\frac{|\a_r|}{2}\bar{\vt}^a_{n(r)}\,,
\end{equation}
and, due to the canonical anti-commutation relations for the fermionic coordinates
\begin{equation}
\{\vt^a_r(\s_r),\l^b_s(\s_s)\}=\d^{ab}\d_{rs}\d(\s_r-\s_s)\,,
\end{equation}
they obey the following anti-commutation rules
\begin{equation}
\{\vt^a_{n(r)},\l^b_{m(s)}\}=\d^{ab}\d_{nm}\d_{rs}\,.
\end{equation}
It is convenient to define a new set of fermionic operators~\cite{sv} 
\begin{equation}
\vt_{n(r)}=\frac{c_{n(r)}}{\sqrt{|\a_r|}}\left[(1+\r_{n(r)}\Pi)b_{n(r)}
+e(\a_r)e(n)(1-\r_{n(r)}\Pi)b_{-n(r)}^{\dg}\right]\,,
\end{equation}
which break the $SO(8)$ symmetry to $SO(4)\times SO(4)$. Here
\begin{equation}
\r_{n(r)}=\r_{-n(r)}=\frac{\o_{n(r)}-|n|}{\m\a_r}\,,\qquad
c_{n(r)}=c_{-n(r)}=\frac{1}{\sqrt{1+\r_{n(r)}^2}}\,.
\end{equation}
These modes satisfy 
\begin{equation}
\{b^a_{n(r)},b^{b\,\dg}_{m(s)}\}=\d^{ab}\d_{nm}\d_{rs}\,.
\end{equation}

\noindent The free string light-cone Hamiltonian is
\begin{equation}
H_{2(r)}=\frac{1}{\a_r}\sum_{n\in\Zop}\o_{n(r)}\bigl(a_{n(r)}^{\dg}a_{n(r)}+b_{n(r)}^{\dg}b_{n(r)}\bigr)\,.
\end{equation}
In the above the zero-point energies cancel between bosons and fermions. 
Since the Hamiltonian only depends on two dimensionful
quantities $\m$ and $\a_r$, $\a'$ and $p^+_r$ should not be
thought of as separate parameters. The vacuum $|v\ra_r$ is defined as
\begin{equation}
a_{n(r)}|v\ra_r=0\,,\qquad b_{n(r)}|v\ra_r=0\,,\quad n\in\Nop\,.
\end{equation}

\noindent The isometries of the pp-wave superalgebra are generated by 
$H$, $P^+$, $P^I$, $J^{+I}$,
$J^{ij}$ and $J^{i'j'}$. The latter two are angular momentum generators of the transverse $SO(4)\times SO(4)$ symmetry
of the pp-wave background. The
32 supersymmetries 
split into kinematical supercharges $Q^+$, $\bar{Q}^+$ and dynamical ones 
$Q^-$, $\bar{Q}^-$. The latter square to the light-cone Hamiltonian and
hence get corrected by interactions.
A subset of the superalgebra that will be of
importance throughout this paper
is~\cite{m}
\begin{equation}\label{comm}
\begin{split}
[H,P^I] &= -i\m^2J^{+I}\,,\qquad [H,Q^+]=-\m\Pi Q^+\,,\\
\{Q^-_{\da},\bar{Q}^-_{\db}\} &=
2\d_{\dot{a}\dot{b}}H-i\m\bigl(\g_{ij}\Pi\bigr)_{\dot{a}\dot{b}}J^{ij}
+i\m\bigl(\g_{i'j'}\Pi\bigr)_{\dot{a}\dot{b}}J^{i'j'}\,.
\end{split}
\end{equation}
The dynamical supercharges can be expressed in terms of world-sheet fields 
as
\begin{equation}\label{qfield}
Q^-_{(r)} 
=\sqrt{\frac{2}{\a'}}\int_0^{2\pi|\a_r|}d\s_r\,\left[2\pi\a'e(\a_r)p_r\g\l_r-
ix'_r\g\bar{\l}_r-i\m x_r\g\Pi\l_r\right]
\end{equation}
and $\bar{Q}^-_{(r)}=e(\a_r)\bigl[Q_{(r)}^-\bigr]^{\dg}$.
Expanding in modes one finds
\begin{align}\label{q-mode}
Q^-_{(r)}& = \frac{e(\a_r)}{\sqrt{|\a_r|}}\g
\Bigl(\sqrt{\m}\left[a_{0(r)}(1+e(\a_r)\Pi)+a_{0(r)}^{\dg}(1-e(\a_r)\Pi)\right]\l_{0(r)}
\nonumber\\
&+\sum_{n\neq 0}\sqrt{|n|}\left[a_{n(r)}P_{n(r)}^{-1}b_{n(r)}^{\dg}
+e(\a_rn)a_{n(r)}^{\dg}P_{n(r)}b_{-n(r)}\right]\Bigr)\,,
\end{align}
where
\begin{equation}
P_{n(r)}=\frac{1}{\sqrt{1-\rho_{n(r)}^2}}(1-\rho_{n(r)}\Pi)\,.
\end{equation}
\section{Interacting string theory}\label{sec3}
In interacting light-cone string field theory some generators  of the
supersymmetry algebra, such as $J$, $P$ and $Q^+$, are quadratic in fields. These kinematical generators correspond to 
symmetries which are manifest in the theory and do not receive $g_{\text{s}}$
corrections. On the other hand, dynamical generators such as $H$, $Q^-$ and $\bar{Q}^-$, depend on the interactions
and hence involve higher powers of fields. 
For example $H$, the full Hamiltonian of the interacting theory, has an expansion in string coupling
\begin{equation}
H=H_2+g_{\text{s}}H_3+\cdots\,,
\end{equation}
where $H_3$ creates or destroys a single string, while the $H_2$ conserves 
the number of strings. 

\noindent Despite such corrections to the generators, the supersymmetry 
algebra has to be satisfied order by order in $g_{\text{s}}$. This consistency condition gives rise to two
types of constraints: the kinematic constraints arise
from demanding that (anti)commutation relations of kinematical with dynamical
generators are realized at order ${\mc O}(g_{\text{s}})$; the dynamic constraints arise from the 
(anti)commutation relations of dynamical generators alone. The former are relatively easy to satisfy, and
it is the latter relations that severely constrain the
form of the dynamical generators and will be used to determine the interaction vertex below.
In this section we review the exponential part of the cubic vertex which enforces
the kinematic constraints.\footnote{Apart from the differences mentioned in the introduction,
our expression for the exponential part of the vertex agrees with the construction first presented in~\cite{sv}.}
In the following section we construct the prefactor of the interaction vertex
by imposing the dynamic constraints.

\noindent The bosonic contribution $|E_a\ra$ to the exponential part of the three-string interaction
vertex has to satisfy the kinematic constraints \cite{gs,gsb}
\begin{equation}\label{kinb}
\sum_{r=1}^3p_r(\s_r)|E_a\ra=0\,,\qquad \sum_{r=1}^3e(\a_r)x_r(\s_r)|E_a\ra=0\,.
\end{equation}
These are the same as in flat space and arise from the commutation relations of $H$ with
$P^I$ ({\it cf}. equation \eqref{comm}) and $J^{+I}$.
They guarantee momentum conservation and continuity of the string worldsheet in the interaction. 
The coordinates of the three strings are parameterized by
\begin{align}
\s_1 & =\s \qquad\quad\qquad -\pi\a_1\le\s\le\pi\a_1\,, \nonumber \\
\s_2 & =\begin{cases} \s-\pi\a_1 & \quad\pi\a_1\le\s\le\pi(\a_1+\a_2)\,, \\
\s+\pi\a_1 & \quad-\pi(\a_1+\a_2)\le\s\le-\pi\a_1\,, \end{cases} \\
\s_3 & =-\s \qquad\quad\qquad -\pi(\a_1+\a_2)\le\s\le \pi(\a_1+\a_2)
\end{align}
and $\a_1+\a_2+\a_3=0$, $\a_3<0$.
The solution to the constraints in equation \eqref{kinb} can be constructed as a functional
integral and is given by \cite{sv}
\begin{equation}\label{bv}
|E_a\ra\sim\exp\left(\frac{1}{2}
\sum_{r,s=1}^3\sum_{m,n\in\Zop}a^{\dg}_{m(r)}\bar{N}^{rs}_{mn}a^{\dg}_{n(s)}
\right)|0\ra_{123}\,,
\end{equation}
where $|0\ra_{123}=|0\ra_1\otimes|0\ra_2\otimes|0\ra_3$ is annihilated by $a_{n(r)}$, $n\in\Zop$.
The determinant factor coming from the functional integral will be cancelled by the
fermionic determinant except for the zero-mode part which is
$\left(\frac{2}{\a'}\frac{2\m}{\pi^3}\frac{\a_1\a_2}{\a_3}\right)^2$.
In the equation above the~non-vanishing elements of the bosonic Neumann
matrices for $m$, $n>0$ are~\cite{sv}
\begin{align}
\label{mn} \bar{N}^{rs}_{mn} & =\d^{rs}\d_{mn}
-2\sqrt{\frac{\o_{m(r)}\o_{n(s)}}{mn}}\left(A^{(r)\,T}\G^{-1}A^{(s)}\right)_{mn}\,,\\
\label{m0} \bar{N}^{rs}_{m0} & =
-\sqrt{2\m\a_s\o_{m(r)}}\e^{st}\a_t\bar{N}^r_m\,,\qquad s\in\{1,2\}\,,\\
\label{00a} \bar{N}^{rs}_{00} & =
(1-4\m\a K)\left(\d^{rs}+\frac{\sqrt{\a_r\a_s}}{\a_3}\right)\,,\qquad  r,s\in\{1,2\}\,,\\
\label{00b} \bar{N}^{r3}_{00} & =
-\sqrt{-\frac{\a_r}{\a_3}}\,,\qquad
r\in\{1,2\}\,.
\end{align}
Here 
\begin{equation}
\a\equiv\a_1\a_2\a_3
\end{equation}
and
\begin{equation}
\G\equiv\sum_{r=1}^3 A^{(r)}U_{(r)}A^{(r)\,T}\,,
\end{equation}
where
\begin{equation}
U_{(r)}\equiv C^{-1}\bigl(C_{(r)}-\m\a_r\bigr)\,,\qquad
C_{mn}\equiv m\d_{mn}\,,\quad
\bigl(C_{(r)}\bigr)_{mn}\equiv\o_{m(r)}\d_{mn}\,.
\end{equation}
The matrices $A^{(r)}$ arise from the Fourier transformation of the constraints in
equation \eqref{kinb} and are given in appendix \ref{appE}. Notice that in the Neumann
matrices the inverse of the infinite dimensional matrix $\G$ appears.
In contrast to flat space~\cite{Mand}, in the pp-wave explicit
expressions for the Neumann matrices valid for all $\m$, are not known. 
It is however possible to obtain their large $\m$ behavior using the approach of~\cite{sch} (see also 
appendix~\ref{appE}).
We also define
\begin{equation}
\bar{N}^r\equiv -C^{-1/2}A^{(r)\,T}\G^{-1}B\,,\qquad
K\equiv-\frac{1}{4}B^T\G^{-1}B\,.
\end{equation}
An explicit expression for the vector $B$ is given in appendix~\ref{appE}. The quantities
$\G$, $\bar{N}^r$ and $K$ manifestly reduce to their flat space counterparts,
defined in \cite{gs,gsb}, as $\m\to 0$.
The only non-vanishing matrix elements with negative indices are
$\bar{N}^{rs}_{-m,-n}$. They are related to $\bar{N}^{rs}_{mn}$ via \cite{sv}
\begin{equation}\label{neg}
\bar{N}^{rs}_{-m,-n}=-\left(U_{(r)}\bar{N}^{rs}U_{(s)}\right)_{mn}\,,\qquad m,\,n>0\,.
\end{equation}
A very useful formula relating
$\bar{N}^{rs}_{mn}$ to $\bar{N}^r_m\bar{N}^s_n$ is~\cite{sch,ap}
\begin{align}\label{nnpp}
\bar{N}^{rs}_{mn} & =-(1-4\m\a K)^{-1}\frac{\a}{\a_r\o_{n(s)}+\a_s\o_{m(r)}}
\left[U_{(r)}^{-1}C_{(r)}^{1/2}C\bar{N}^r\right]_m\left[U_{(s)}^{-1}C_{(s)}^{1/2}C\bar{N}^s\right]_n\,.
\end{align}
This factorization theorem can be used to verify~\cite{ap} that $|E_a\ra$ 
satisfies equation~\eqref{kinb}. It will also prove essential throughout the next section.

\noindent Analogously to the bosonic case, the fermionic exponential part of the
interaction vertex has to satisfy \cite{gs,gsb}
\begin{equation}\label{kinf}
\sum_{r=1}^3\l_{(r)}(\s_r)|E_b\ra=0\,,\qquad
\sum_{r=1}^3e(\a_r)\vt_{(r)}(\s_r)|E_b\ra=0\,.
\end{equation}
These constraints arise from the commutation relations of $H$ with $Q^+$ and $\bar{Q}^+$,
{\it cf}.\ equation \eqref{comm}. The solution is \cite{ap}
\begin{equation}\label{fv}
|E_b\ra \sim
\exp\left[\sum_{r,s=1}^3\sum_{m,n=1}^{\infty}b^{\dg}_{-m(r)}Q^{rs}_{mn}b^{\dg}_{n(s)}
-\sqrt{2}\L\sum_{r=1}^3\sum_{m=1}^{\infty}Q^r_mb^{\dg}_{-m(r)}\right]|E_b^0\ra\,,
\end{equation}
where $\L\equiv\a_1\l_{0(2)}-\a_2\l_{0(1)}$ and $|E_b^0\ra$ is the pure zero-mode part of
the fermionic vertex
\begin{equation}\label{zero}
|E^0_b\ra=\prod_{a=1}^8\left[\sum_{r=1}^3\l_{0(r)}^a\right]|0\ra_{123}\,.
\end{equation}
Here $|0\ra_r$ is not the vacuum defined to be annihilated by the $b_{0(r)}$. Rather,
it satisfies $\vt_{0(r)}|0\ra_r=0$ and $H_{2(r)}|0\ra_r=4\m e(\a_r)|0\ra_r$. 
In the limit $\m\to 0$ it coincides with the
flat space state that generates the massless multiplet by acting with $\l_{0(r)}^a$ on it.
The fermionic Neumann matrices can be expressed in terms of the bosonic ones as \cite{ap}
\begin{align}
\label{qmn}
Q^{rs}_{mn} & =e(\a_r)\sqrt{\left|\frac{\a_s}{\a_r}\right|}
\bigl[P_{(r)}^{-1}U_{(r)}C^{1/2}\bar{N}^{rs}C^{-1/2}U_{(s)}P_{(s)}^{-1}\bigr]_{mn}\,,\\
\label{qm}
Q^r_n & =\frac{e(\a_r)}{\sqrt{|\a_r|}}(1-4\m\a
K)^{-1}(1-2\m\a K(1+\Pi))\bigl[P_{(r)}C_{(r)}^{1/2}C^{1/2}\bar{N}^r\bigr]_n\,.
\end{align}
In summary, the part of the cubic contribution to the dynamical generators satisfying the
kinematic constraints is
\begin{equation}
|V\ra\equiv|E_a\ra|E_b\ra\d\left(\sum_{r=1}^3\a_r\right)\,.
\end{equation}
\section{The Complete ${\mc O}(g_{\text{s}})$ Superstring Vertex}\label{sec4}
In the previous section we reviewed the exponential part of the vertex, which solves the kinematic constraints.
The remaining dynamic constraints are much more restrictive and are solved by 
introducing a prefactor \cite{gs,gsb}, polynomial in creation operators, in front of $|V\ra$.\footnote{The prefactors 
are polynomials of ${\mc K},\,{\wt{\mc K}}$ and ${\mc Y}$, defined below,
where ${\mc K},\,{\wt{\mc K}}$ and ${\mc Y}$ are linear in creation operators. Apart from
an important factor of $i$ (see discussion in the introduction),
the oscillator expressions for ${\mc K}$ and ${\wt{\mc K}}$ were first
derived in~\cite{sv2}. The corresponding expression for ${\mc Y}$ was first derived
in~\cite{ap}.}
Within the functional formalism, the prefactor can be re-interpreted as an insertions of local operators
at the interaction point~\cite{Mandsusy,gs} (see also appendix~\ref{appf}).
In this section we present expressions for the dynamical generators
and prove that they satisfy the superalgebra up to order ${\mc O}(g_{\text{s}})$.
\subsection{The Superalgebra and the Constituents of the Prefactors}\label{sec41}
Define the linear combinations of the free supercharges ($\eta=e^{i\pi/4}$)
\begin{equation}
\sqrt{2}\eta\,Q\equiv Q^-+i\bar{Q}^-\,,\qquad\text{and}\qquad
\sqrt{2}\bar{\eta}\,\wt{Q}=Q^--i\bar{Q}^-
\end{equation}
which satisfy
\begin{equation}
\begin{split}
\{Q_{\dot{a}},\wt{Q}_{\dot{b}}\} & = 
-\m\bigl(\g_{ij}\Pi\bigr)_{\dot{a}\dot{b}}J^{ij}
+\m\bigl(\g_{i'j'}\Pi\bigr)_{\dot{a}\dot{b}}J^{i'j'}
\,,\\
\{Q_{\dot{a}},Q_{\dot{b}}\}& = \{\wt{Q}_{\dot{a}},\wt{Q}_{\dot{b}}\}
=2\d_{\dot{a}\dot{b}}H\,.
\end{split}
\end{equation}
Since $J^{ij}$ and $J^{i'j'}$ are not corrected by the interaction, 
it follows that at order ${\mc O}(g_{\text{s}})$ the dynamical generators have to satisfy
\begin{eqnarray}
\label{dyn1}
\sum_{r=1}^3Q_{\dot{a}(r)}|Q_{3\,\dot{b}}\ra+\sum_{r=1}^3Q_{\dot{b}(r)}|Q_{3\,\dot{a}}\ra
& =& 2|H_3\ra\d_{\dot{a}\dot{b}}\,,\\
\label{dyn2}
\sum_{r=1}^3\wt{Q}_{\dot{a}(r)}|\wt{Q}_{3\,\dot{b}}\ra+\sum_{r=1}^3\wt{Q}_{\dot{b}(r)}|\wt{Q}_{3\,\dot{a}}\ra
&=& 2|H_3\ra\d_{\dot{a}\dot{b}}\,,\\
\label{dyn3}
\sum_{r=1}^3Q_{\dot{a}(r)}|\wt{Q}_{3\,\dot{b}}\ra+\sum_{r=1}^3\wt{Q}_{\dot{b}(r)}|Q_{3\,\dot{a}}\ra  &=&0\,.
\end{eqnarray}
In order to derive equations that determine the full expressions for the dynamical
generators one has to compute (anti)commutators of $Q_{\da(r)}$ and $\wt{Q}_{\da(r)}$ with the
constituents of the prefactors.
Moreover, the action of the supercharges on
$|V\ra$ has to be known in terms of these constituents. 
Here the factorization theorem~\eqref{nnpp} for the bosonic Neumann matrices
and the relation between the bosonic and fermionic Neumann matrices given in equations~\eqref{qmn} and~\eqref{qm}
prove to be essential.

\noindent
So as not to destroy the constraints in equation \eqref{kinb}, 
the bosonic part of the prefactors (that we will collectively denote by ${\mc P}$) has to
satisfy~\cite{gs,gsb}
\begin{equation}\label{bospre}
\bigl[\,\sum_{r=1}^3p_r(\s_r),{\mc P}\bigr]=0
=\bigl[\,\sum_{r=1}^3e(\a_r)x_r(\s_r),\mc{P}\bigr]\,.
\end{equation}
Making an ansatz linear in creation oscillators and taking the Fourier transform of the above
equations one finds that negative and non-negative modes decouple from each other. The
resulting constraint equations on the infinite component vectors appearing in the ansatz
can be solved using flat space identities~\cite{gs}; for details, see \cite{sv2,ap}.
The solution of these constraints involving non-negatively moded bosonic creation oscillators is \cite{sv2}
\begin{equation}\label{k+mode}
{\mc K}_0+{\mc K}_+\equiv
\mathbb{P}-i\m\frac{\a}{\a'}\mathbb{R}
+\sum_{r=1}^3\sum_{n=1}^{\infty}F_{n(r)}a^{\dg}_{n(r)}\,,
\end{equation}
where
\begin{equation}
\mathbb{P}\equiv\a_1p_{0(2)}-\a_2p_{0(1)}\,,\qquad
\a_3\mathbb{R}\equiv x_{0(1)}-x_{0(2)}\,,\qquad [\mathbb{R},\mathbb{P}]=i
\end{equation}
and in terms of the zero-mode creation oscillators
\begin{equation}
\sum_{r=1}^3F_{0(r)}a_{0(r)}^{\dg}\equiv\mathbb{P}-i\m\frac{\a}{\a'}\mathbb{R}=
\sqrt{\frac{2}{\a'}}\sqrt{\m\a_1\a_2}
\bigl(\sqrt{\a_1}a_{0(2)}^{\dg}-\sqrt{\a_2}a_{0(1)}^{\dg}\bigr)\,.
\end{equation}
The explicit expression for $F_{n(r)}$ is given in appendix \ref{appB}, equation \eqref{f+}.
The solution involving negatively moded creation oscillators is
\begin{equation}\label{k-mode}
{\mc K}_-\equiv\sum_{r=1}^3\sum_{n=1}^{\infty}F_{-n(r)}a^{\dg}_{-n(r)}\,,
\end{equation}
where
\begin{equation}\label{negf}
F_{-n(r)}=iU_{n(r)}F_{n(r)}\,.
\end{equation}

\noindent The fermionic parts of the prefactor have to satisfy the conditions
\begin{equation}\label{ferpre}
\bigl\{\,\sum_{r=1}^3\l_r(\s),{\mc P}\bigr\}=0=
\bigl\{\,\sum_{r=1}^3e(\a_r)\vt_r(\s),\mc{P}\bigr\}
\end{equation}
that are solved in a similar way to the bosonic case.
The solution involving non-negatively moded creation oscillators is
\begin{equation}\label{ymode}
{\mc Y}\equiv\sqrt{\frac{2}{\a'}}\L
+\sum_{r=1}^3\sum_{n=1}^{\infty}G_{n(r)}b^{\dg}_{n(r)}\,,
\end{equation}
where the infinite component vector $G_{n(r)}$ can be expressed in terms of the
bosonic one $F_{n(r)}$ \cite{ap} (see appendix \ref{appB}, equation \eqref{gf}).
As in flat space \cite{gs,gsb}, it turns out that the prefactors
do not involve negatively moded fermionic creation oscillators.

\noindent
Below we present the results necessary to verify equations
\eqref{dyn1} and \eqref{dyn2}, given the ansatz \eqref{h}-\eqref{tq} for the cubic vertex and 
dynamical supercharges.
We need
\begin{equation}\label{qk}
\sqrt{2}\eta\sum_{r=1}^3[Q_{(r)},\wt{\mc K}]\,|V\ra=
\sqrt{2}\bar{\eta}\sum_{r=1}^3[\wt{Q}_{(r)},{\mc K}]\,|V\ra=
\m\g(1+\Pi){\mc Y}|V\ra\,,
\end{equation}
where
\begin{equation}
{\mc K}\equiv{\mc K}_0+{\mc K}_++{\mc K}_-\,,\qquad
\wt{\mc K}\equiv{\mc K}_0+{\mc K}_+-{\mc K}_-
\end{equation}
and
\begin{equation}\label{qy}
\begin{split}
\sqrt{2}\eta\sum_{r=1}^3\{Q_{(r)},{\mc Y}\}\wt{\mc K}^I|V\ra 
& =
i\g^J(1-2\m\a K(1-\Pi)){\mc K}^J\wt{\mc K}^I|V\ra-i\m\frac{\a}{\a'}\g^I(1-\Pi)|V\ra\,,\\
\sqrt{2}\bar{\eta}\sum_{r=1}^3\{\wt{Q}_{(r)},{\mc Y}\}{\mc 
K}^I|V\ra & =
-i\g^J(1-2\m\a K(1-\Pi))\wt{\mc K}^J{\mc K}^I|V\ra+i\m\frac{\a}{\a'}\g^I(1-\Pi)|V\ra\,.
\end{split}
\end{equation}
Notice, that the above identities are only valid when both sides of the equation act on $|V\ra$.
The action of the supercharges on $|V\ra$ is 
\begin{align}
\label{qv}
\sqrt{2}\eta\sum_{r=1}^3Q_{(r)}|V\ra & =
-\frac{\a'}{\a}{\mc K}\g(1-2\m\a K(1+\Pi)){\mc Y}|V\ra\,,\\
\label{tqv}
\sqrt{2}\bar{\eta}\sum_{r=1}^3\wt{Q}_{(r)}|V\ra & =
-\frac{\a'}{\a}\wt{\mc K}\g(1-2\m\a K(1+\Pi)){\mc Y}|V\ra\,.
\end{align}
The proof of equations \eqref{qk}-\eqref{tqv} is given in appendix \ref{appB}. 
\subsection{The dynamical generators at order ${\mc O}(g_{\text{s}})$}\label{sec42}
The results of the previous subsection motivate the following ansatz for the explicit form
of the dynamical supercharges and the three-string interaction vertex
\begin{align}
\label{h}
|H_3\ra & =
\left((1-4\m\a K)\wt{\mc K}^I{\mc K}^J-\m\frac{\a}{\a'}\d^{IJ}\right)v_{IJ}(Y)|V\ra\,,\\
\label{q}
|Q_{3\,\dot{a}}\ra & = (1-4\m\a K)^{1/2}\wt{\mc K}^Is_{\dot{a}}^I(Y)|V\ra\,,\\
\label{tq}
|\wt{Q}_{3\,\dot{a}}\ra & = (1-4\m\a K)^{1/2}{\mc K}^I\tilde{s}_{\dot{a}}^I(Y)|V\ra\,.
\end{align}
Here
\begin{equation}\label{y}
Y\equiv(1-4\m\a K)^{-1/2}(1-2\m\a K(1+\Pi)){\mc Y}\,.
\end{equation}
Substituting this
into equations \eqref{dyn1} and \eqref{dyn2} and using equations \eqref{qk}-\eqref{tqv}, 
we get the following equations for
$v^{IJ}$, $s_{\da}^I$ and $\tilde{s}_{\da}^I$\,\footnote{Here $(\da\db)$ denotes symmetrization in $\da$, $\db$.}
\begin{equation}\label{flat}
\d_{\dot{a}\dot{b}}v^{IJ} =
\frac{i}{\sqrt{2}}\frac{\a'}{\a}\g^J_{a(\dot{a}}D^as^I_{\dot{b})}\,,\qquad
\d_{\dot{a}\dot{b}}v^{IJ} =
-\frac{i}{\sqrt{2}}\frac{\a'}{\a}\g^I_{a(\dot{a}}\bar{D}^a
\tilde{s}^J_{\dot{b})}\,,
\end{equation}
which originate from terms proportional to $\wt{\mc K}_I{\mc K}_J$ and are identical to the flat
space equations of \cite{gsb}.
Two additional equations, arising from terms proportional to $\m\d_{IJ}$ are
\begin{equation}\label{ppwave1}
\begin{split}
-\d_{\dot{a}\dot{b}}v^{II} & =\frac{i}{\sqrt{2}}\frac{\a'}{\a}
\g^I_{a(\dot{a}}\Bigl(D^a+i\bigl[\Pi\bar{D}\bigr]^a\Bigr)s^I_{\dot{b})}\,,\\
-\d_{\dot{a}\dot{b}}v^{II} & = -\frac{i}{\sqrt{2}}\frac{\a'}{\a}
\g^I_{a(\dot{a}}\Bigl(\bar{D}^a-i\bigl[\Pi D\bigr]^a\Bigr)\tilde{s}^I_{\dot{b})}\,.
\end{split}
\end{equation}
As in flat space \cite{gsb} we define
\begin{equation}
D^a\equiv\eta Y^a+\bar{\eta}\frac{\a}{\a'}\frac{\p}{\p Y_a}\,,\qquad
\bar{D}^a\equiv\bar{\eta} Y^a+\eta\frac{\a}{\a'}\frac{\p}{\p Y_a}\,.
\end{equation}
Recall first the solution of the flat space equations \eqref{flat} \cite{gsb}. One introduces the following functions
of $Y^a$
\begin{align}
w^{IJ} & = \d^{IJ} +\left(\frac{\a'}{\a}\right)^2\frac{1}{4!}t^{IJ}_{abcd}Y^aY^bY^cY^d
+\left(\frac{\a'}{\a}\right)^4\frac{1}{8!}\d^{IJ}\e_{abcdefgh}Y^a\cdots Y^h\,,\\
iy^{IJ} & = \frac{\a'}{\a}\frac{1}{2!}\g^{IJ}_{ab}Y^aY^b+\left(\frac{\a'}{\a}\right)^3\frac{1}{2\cdot 6!}
\g^{IJ}_{ab}{\e^{ab}}_{cdefgh}Y^c\cdots Y^h\,,\\
\frac{1}{2}s^I_{1\,\dot{a}} & = 
\g^I_{a\dot{a}}Y^a+\left(\frac{\a'}{\a}\right)^2\frac{1}{6!}
u^I_{abc\dot{a}}{\e^{abc}}_{defgh}Y^d\cdots Y^h\,,\\
\frac{1}{2}s^I_{2\,\dot{a}} & = 
-\frac{\a'}{\a}\frac{1}{3!}u^I_{abc\dot{a}}Y^aY^bY^c+
\left(\frac{\a'}{\a}\right)^3\frac{1}{7!}\g^I_{a\dot{a}}{\e^{a}}_{bcdefgh}Y^b\cdots Y^h\,.
\end{align}
Here
\begin{equation}
t^{IJ}_{abcd}\equiv\g^{IK}_{[ab}\g^{JK}_{cd]}\,,\qquad u^I_{abc\dot{a}}\equiv -\g^{IJ}_{[ab}\g^J_{c]\dot{a}}\,.
\end{equation}
Notice that $t^{IJ}_{abcd}$ is traceless and symmetric in $I$, $J$, hence $w^{IJ}$
is a symmetric tensor of $SO(8)$, whereas $y^{IJ}$ is antisymmetric. 
The solution of equations \eqref{flat} is~\cite{gsb}
\begin{equation}\label{sol}
v^{IJ}\equiv w^{IJ}+y^{IJ}\,,\qquad
s^I_{\dot{a}}\equiv-\frac{2}{\a'}\frac{i}{\sqrt{2}}
\bigl(\eta s^I_{1\,\dot{a}}+\bar{\eta}s^I_{2\,\dot{a}}\bigr)\,,\qquad
\tilde{s}^I_{\dot{a}}\equiv \frac{2}{\a'}\frac{i}{\sqrt{2}}
\bigl(\bar{\eta} s^I_{1\,\dot{a}}+\eta s^I_{2\,\dot{a}}\bigr)\,.
\end{equation}

\noindent Next consider the additional equations \eqref{ppwave1}. Using the flat space
solution, these can be rewritten as
\begin{equation}
0=\g^I_{a(\dot{a}}\bigl[\Pi\bar{D}\bigr]^as^I_{\dot{b})}\,\qquad
0=\g^I_{a(\dot{a}}\bigl[\Pi D\bigr]^a\tilde{s}^I_{\dot{b})}\,.
\end{equation}
We prove in appendix \ref{appD} that these equations are also 
satisfied by equation~\eqref{sol}.

\noindent The proof of equation \eqref{dyn3} is more involved and provides an important consistency
check of the ansatz~\eqref{h}--\eqref{tq}. We show in appendix \ref{appC} that it leads to the equations
\begin{align}
\label{qq1}
&\d^{IJ}m_{\da\db}-\frac{1}{\sqrt{2}}\frac{\a'}{\a}\g^{(I}_{a\da}D^a\tilde{s}^{J)}_{\db}=0\,,\\
&\d^{IJ}m_{\da\db}-\frac{1}{\sqrt{2}}\frac{\a'}{\a}\g^{(I}_{a\db}\bar{D}^as^{J)}_{\da}=0\,,\\
\label{qq3}
&\sqrt{2}\bigl(\g^I_{a\da}\eta\tilde{s}^I_{\db}-\g^{I}_{a\db}\bar{\eta}s^I_{\da}\bigr)-4im_{\da\db}Y_a=0\,,\\
\label{ppwave2}
&\bigl(\g^I_{a\da}\bar{D}_b\tilde{s}^I_{\db}+\g^I_{a\db}D_bs^I_{\da}\bigr)(1-\Pi)^{ab}=0\,.
\end{align}
Here
\begin{align}
m_{\dot{a}\dot{b}} & =\d_{\dot{a}\dot{b}}+\frac{i}{4}\frac{\a'}{2\a}\g^{IJ}_{\dot{a}\dot{b}}\g^{IJ}_{ab}Y^aY^b
-\frac{1}{4\cdot 4!}\left(\frac{\a'}{2\a}\right)^2\g^{IJKL}_{\dot{a}\dot{b}}t^{IJKL}_{abcd}Y^aY^bY^cY^d
\nonumber\\
&-\frac{i}{6!}\left(\frac{\a'}{2\a}\right)^3\g^{IJ}_{\dot{a}\dot{b}}\g^{IJ}_{ab}{\e^{ab}}_{cdefgh}Y^c\cdots
Y^h-\frac{2}{7!}\left(\frac{\a'}{2\a}\right)^4\d_{\dot{a}\dot{b}}\e_{abcdefgh}Y^a\cdots Y^h
\end{align}
and
\begin{equation}
t^{IJKL}_{abcd}\equiv\g^{[IJ}_{[ab}\g^{KL]}_{cd]}\,.
\end{equation}
The first three equations are identical to those in flat space\footnote{In appendix~\ref{appC} we correct some 
minor typos present in~\cite{gsb}.} and were proven in \cite{gsb}.
The additional equation \eqref{ppwave2} is proved in appendix \ref{appD}. 

\noindent We have not yet fixed the overall normalisation of the
dynamical generators which can depend on $\m$ and the $\a_r$'s. 
This is more difficult to do than in flat space since there is no $J^{-I}$ generator in the 
pp-wave background. 
A comparison with a
supergravity calculation fixes the normalisation for small $\m$ to be $\sim(\a'\m^2)/(\a_3^4)$ \cite{kim}.
However, this does not
completely fix the normalisation; one may still multiply by a function of
$\m$ and $\a_r$ that goes to one as $\m\to 0$. On the other hand, the definition of $Y$ in
equation~\eqref{y} and the fact that the terms $\wt{\mc K}^I{\mc K}^J$ and $\m\d^{IJ}$ in equation \eqref{h} 
are multiplied by different powers of $1-4 \m\a K$ are fixed by demanding that one recovers the flat space 
equations~\eqref{flat}, \eqref{qq1}-\eqref{qq3}.
In order to obtain the supergravity expressions for the dynamical
generators from equations \eqref{h}--\eqref{tq}, one should set $K$ to zero, since it originates from  
massive string modes. Together with
$[\mathbb{R}^I,\mathbb{P}^J]=i\d^{IJ}$, one can check that the supergravity vertex obtained in this way
agrees with the supergravity vertex presented in section 4 of~\cite{sv}.

\noindent We would like to stress that the nontrivial relation between $Y$ and ${\mc Y}$ in equation \eqref{y}, as 
well as the part in the cubic interaction vertex proportional to $\m\d^{IJ}$, were not present in 
\cite{sv,sv2} and played an essential role in the above proof of the superalgebra in the interacting string field 
theory. 
In~\cite{sv}, functional expressions for the constituents of the prefactors were used to argue that the vertex 
of~\cite{sv} satisfied the superalgebra. 
It is well known~\cite{gs, gsb} that these functional expressions do not in general agree with the oscillator basis 
expressions given in equations \eqref{k+mode}, \eqref{k-mode} and \eqref{ymode}.
As explained in appendix \ref{appf}, it would appear that this subtlety is the origin for 
the absence of the $\m\d^{IJ}$ term in~\cite{sv}. In flat space the functional expression agrees with the one in the 
oscillator basis only when acting directly on $|V\ra$~\cite{gs,gsb}. In the pp-wave however, they agree only up to 
non-trivial functions of $\m$~\cite{ap} (see also appendix~\ref{appf}). This leads to equation \eqref{y}. 
In summary the
oscillator expression~\eqref{h}, which is used in the following section to compute three-point functions,
has been shown explicitly to satisfy the superalgebra at ${\mc O}(g_{\text{s}})$.
\section{3-string amplitudes}\label{sec5}
In this section we use $H_3$ to compute three-string amplitudes. 
For three general string states $|\Phi_i\ra$ in the number-basis representation
the amplitude is
\begin{equation}
g_{\text{s}}\la\Phi_3|H_3|\Phi_1\ra|\Phi_2\ra =
g_{\text{s}}\la\Phi_3|\la\Phi_2|\la\Phi_1|H_3\ra\,.
\end{equation}
We restrict ourselves to the case, where either all states have bosonic excitations or fermionic ones. 
We compare our expressions to gauge theory results in the BMN limit. 
Recall that the full interacting Hamiltonian $H/\m$ in string theory is identified with the operator 
$\D-J$ in field theory~\cite{bmn,gross,verlinde}. A comparison of matrix elements of the two operators
requires an identification of the free string basis used in string theory with a dual basis in field theory. 
At the planar level, multi-trace BMN operators are identified with 
multi-string states~\cite{bmn}. However, non-planar corrections lead to a mixing of single and multi trace BMN 
operators~\cite{jan2,freedman2} and this effect has to be taken into account when comparing
matrix elements of $H_3$ to field theory. The dual basis was proposed~\cite{vv,vsv,gomis} to be the one, in which 
the matrix of two-point functions of redefined BMN operators ({\it i.e.} now mixtures of single and multi-trace)
is diagonal in free field theory. As emphasized in~\cite{vsv,gomis}, the required basis transformation is not 
unique.\footnote{We are grateful to Jan Plefka for discussions on this point.}
\subsection{Amplitudes of bosonic states}\label{sec51}
To facilitate the comparison with field theory, we change our oscillator basis to the BMN basis.
For $n>0$ the transformation is
\begin{equation}
\sqrt{2}a_n=\a_n+\a_{-n}\,,\qquad i\sqrt{2}a_{-n}=\a_n-\a_{-n}\,,\qquad a_0=\a_0\,.
\end{equation}
Using equation \eqref{negf} we have 
\begin{equation}
{\mc K}=\sum_{r=1}^3\sum_{n\in\Zop}{\mc K}_{n(r)}\a_{n(r)}^{\dg}\,,\qquad
\wt{{\mc K}}=\sum_{r=1}^3\sum_{n\in\Zop}\wt{{\mc K}}_{n(r)}\a_{n(r)}^{\dg}\,,
\end{equation}
where
\begin{equation}
{\mc K}_{n(r)}=\begin{cases}
F_{0(r)} & ,\,n=0 \\ \frac{1}{\sqrt{2}}F_{|n|(r)}\bigl(1-U_{n(r)}\bigr) &,\,n\neq0
\end{cases}
\,,\qquad
\wt{{\mc K}}_{n(r)}=\begin{cases}
F_{0(r)} & ,\,n=0 \\ \frac{1}{\sqrt{2}}F_{|n|(r)}\bigl(1+U_{n(r)}\bigr) &,\,n\neq0
\end{cases}\,.
\end{equation}
In this basis the Neumann matrices are 
\begin{equation}
\wt{N}^{rs}_{mn}=
\begin{cases}
\frac{1}{2}\bar{N}^{rs}_{|m||n|}\bigl(1+U_{m(r)}U_{n(s)}\bigr) & \,,m\,,n\neq0 \\
\frac{1}{\sqrt{2}}\bar{N}^{rs}_{|m|0} & \,,m\neq0 \\
\bar{N}^{rs}_{00}\,.
\end{cases}
\end{equation}
\noindent We compute amplitudes involving the string states
\begin{equation}\label{bstates}
|v\ra_r\,,\qquad
|0_i\ra_r\equiv\a_{0(r)}^{i\,\dg}|v\ra_r\,,\qquad
|n_{i,j}\ra_r\equiv\a_{n(r)}^{i\,\dg}\a_{-n(r)}^{j\,\dg}|v\ra_r\,.
\end{equation}
The fermionic contribution to these amplitudes is simple to determine.
The pp-wave vacua ${}_r\la v|$ are related to ${}_r\la 0|$ via
\begin{equation}\label{vo}
{}_r\la v|={}_r\la 0|\left(\frac{\a_r}{2}\right)^2\prod_{a=5}^8\vt^a_{0(r)}\,,\quad r\in\{1,2\}\,,\qquad
{}_3\la v|=-{}_3\la 
0|\left(\frac{\a_3}{2}\right)^2\prod_{a=1}^4\vt^a_{0(r)}\,.
\end{equation}
Eight of the zero-modes in equation \eqref{vo}, 
namely $\vt_{0(3)}^a$, $a=1,\ldots,4$ and, say, $\vt_{0(2)}^a$, $a=5,\ldots,8$ are saturated by $|E_b^0\ra$, so, to 
give a non-zero contribution the remaining four zero-modes must be contracted with the ${\mc O}(Y^4)$ term in 
$v_{IJ}(Y)$. Hence, the fermionic contribution is
\begin{equation}
-\left(\frac{\a_3}{2}\right)^4(1-4\m\a K)^{-2}\Pi_{IJ}\,,
\end{equation}
where $\Pi^{IJ}\equiv t^{IJ}_{5678}=\text{diag}({\bf 1}_4\,,-{\bf 1}_4)$; 
the factor $(1-4\m\a K)^{-2}$ is due to equation~\eqref{y}.
The bosonic zero-mode determinant can be computed using results in appendix B of~\cite{sv}
\begin{equation}
\left(\frac{2}{\a'}\frac{2\m}{\pi^3}\frac{\a_1\a_2}{\a_3}\right)^2\,.
\end{equation}
As discussed at the end of
section \ref{sec4} the overall normalisation of the cubic interaction vertex is
still not completely fixed. We will take it to be
\begin{equation}\label{norm}
\frac{\pi^7{\a'}^3}{\a^2}(1-4\m\a K)^2
\end{equation}
by comparing a three-string amplitude ({\it e.g.} equation \eqref{00n1}) in the limit $\m\to\infty$ to a known
field theory result. Strictly speaking, from the comparison with supergravity and field theory we
only know that equation \eqref{norm} is correct for $\m=0$ and $\m\to\infty$. Since the
function $(1-4\m\a K)$ plays such a prominent role in pp-wave string field theory, we
conjecture that it is the correct choice for all $\m$. Combining the various contributions
we conclude that for amplitudes of string states build out of bosonic oscillators acting on the $|v\ra_r$, the
interaction vertex is\footnote{For general amplitudes it is
$g_{\text{s}}|H_3\ra=16\pi g_{\text{s}}\a'\m^2\a_3^{-4}(1-4\m\a K)^2
\left((1-4\m\a K){\mc K}^I\wt{\mc K}^J-\m\frac{\a}{\a'}\d^{IJ}\right)v_{IJ}(Y)|V\ra$\,.}
\begin{equation}
g_{\text{s}}|H_3\ra=-\pi g_{\text{s}}\a'\m^2(1-4\m\a K){\mc K}^I\wt{\mc K}^J\Pi_{IJ}|E_a\ra
\d\left(\sum_{r=1}^3\a_r\right)\,.
\end{equation}
In particular, the term proportional to $\m\d^{IJ}$ does not contribute, since $\Pi_{IJ}$ is traceless. 
Using the factorization theorem \eqref{nnpp} this can be 
further simplified to~\cite{vsv,Lee:2002vz}
\begin{equation}\label{heff}
g_{\text{s}}|H_3\ra 
=\frac{1}{2}g_2\b(\b+1)\sum_{r=1}^3\sum_{n\in\Zop}\frac{\o_{n(r)}}{\a_r}\a^{I\,\dg}_{n(r)}
\a^J_{-n(r)}\Pi_{IJ}|E_a\ra|\a_3|\d\left(\sum_{r=1}^3\a_r\right)\,.
\end{equation}
Recall that $g_{\text{s}}=g_2/(4\pi\m^2\a_3^2)$. 
Here $\b=\a_1/\a_3$ is related to $r=J_1/J$, the ratio of R-charges 
conventionally used in SYM, by $\b=-r$.
For amplitudes involving two zero-mode oscillators and one massive string state
we find\footnote{To avoid clutter of notation we suppress the index $r$ of the $r$th string and the factor of 
$|\a_3|\d\bigl(\sum_r\a_r\bigr)$.}
\begin{align}
g_{\text{s}}\la n_{1,2}|H_3|0_1\ra|0_2\ra & =
-g_2\m\a_1\a_2(-\b(\b+1))^{3/2}\left[CU_{(3)}^{-1}C_{(3)}\right]_n
\left(\bar{N}^3_{|n|}\right)^2\nonumber\\
\label{00n1}
&=-\m g_2\l'\sqrt{-\b(\b+1)}\frac{\sin^2n\pi\b}{2\pi^2}
\left[1-\frac{1}{4}(3-16a_R)\l'n^2+{\mc O}(\l^{\prime\,2})\right]\,,\\
g_{\text{s}}\la n_{1,1}|H_3|0_1\ra|0_1\ra & =
-2g_2\m\a_1\a_2(-\b(\b+1))^{3/2}\left[CU_{(3)}^{-1}C_{(3)}\right]_n\left(\bar{N}^3_{|n|}\right)^2
\left[1+\frac{1}{4\m\a_3}\left[CU_{(3)}^{-1}\right]_n\right]\nonumber\\
\label{00n2}
& =-\m g_2\l'\sqrt{-\b(\b+1)}\frac{\sin^2n\pi\b}{\pi^2}
\left[1-\frac{1}{4}(7-16a_R)\l'n^2+{\mc O}(\l^{\prime\,2})\right]\,.
\end{align}
The leading order result given in equation \eqref{00n1} has also been computed in
\cite{vsv,gomis} and is in agreement with field theory~\cite{jan2,freedman2}. The subleading terms are
predictions for the 2-loop results in field theory from which one should be able to
fix the yet unknown constant $a_R$ ({\it cf.} appendix~\ref{appE}). An amplitude of two
massive string states and the pp-wave vacuum is
\begin{align}
g_{\text{s}}\la n_{1,2}|H_3|m_{1,2}\ra|v\ra & =
g_2\frac{\b+1}{4\a_3^2}\bigl(\a_1\o_{n(3)}+\a_3\o_{m(1)}\bigr)
\left(\bar{N}^{13}_{|m||n|}\right)^2\bigl(1-U_{m(1)}^2U_{n(3)}^2\bigr)\nonumber\\
\label{m0n}
&=\m g_2\l'(\b+1)\frac{\sin^2n\pi\b}{2\pi^2}+{\mc O}(\l^{\prime\,2})\,.
\end{align}
This amplitude has also been computed in \cite{vsv,gomis} and agrees with
the field theory result~\cite{jan2,freedman2}. Finally, consider amplitudes
involving only zero-mode oscillators. From equation \eqref{heff} it follows, that
all amplitudes with, say, $l$ oscillators for the first string, $m$ oscillators for the
second string, $n$ oscillators for the third string and $l+m=n$ vanish for all $\m$.
This is in agreement with the proposal of \cite{vv,vsv,gomis}, since 
the operators
in field theory dual to single and double string states at order ${\mc O}(g_2)$ are
constructed in such a way that they do not mix in free field theory. Since the operator 
dual of the single string is protected this is an exact result for all $\l'$, thereby matching the
string theory result. 
\subsection{Amplitudes of fermionic states}\label{sec52}
As in the bosonic case, we change the oscillator basis to the BMN basis ($n>0$)
\begin{equation}
\sqrt{2}b_n=\b_n+\b_{-n}\,,\qquad i\sqrt{2}b_{-n}=\b_n-\b_{-n}\,,\qquad b_0=\b_0\,.
\end{equation}
and compute amplitudes of the states 
\begin{equation}
|v\ra_r\,,\qquad
|0_a\ra_r\equiv\b_{0(r)}^{a\,\dg}|v\ra_r\,,\qquad
|n_{a,b}\ra_r\equiv\b_{n(r)}^{a\,\dg}\b_{-n(r)}^{b\,\dg}|v\ra\,.
\end{equation}
In this case the cubic interaction vertex reduces to 
\begin{equation}
g_{\text{s}}|H_3\ra=-4\m g_2\frac{\b(\b+1)}{\a_3^4}(1-4\m\a K)^2\tr(v)(Y)|E_b\ra
|\a_3|\d\left(\sum_r\a_r\right)\,.
\end{equation}
Due to the trace over $v_{IJ}$ only terms of order ${\mc O}(Y^0)$ and
${\mc O}(Y^8)$ contribute. Note that these amplitudes are identically zero for the vertex 
constructed in~\cite{sv}, because the trace over $v_{IJ}$ comes from the term proportional 
to $\m\d^{IJ}$. 
As an example, we consider 
\begin{equation}\label{feramp}
g_{\text{s}}\la n_{5,6}|H_3|0_7\ra|0_8\ra\,,\qquad
g_{\text{s}}\la n_{5,6}|H_3|m_{7,8}\ra|v\ra\,,
\end{equation}
which are fermionic analogues to the bosonic amplitudes in equations \eqref{00n1} and
\eqref{m0n}. For the above amplitudes to be non-zero it is essential that the fermionic exponential part
of the vertex contains a term linear in zero-modes. 
Counting fermionic zero-modes shows that in this case only
the ${\mc O}(Y^0)$ term in $\tr(v)$ gives a contribution.
We find
\begin{align}
\label{fer1}
g_{\text{s}}\la n_{5,6}|H_3|0_7\ra|0_8\ra & =
-\m\a_1\a_2g_2\bigl(-\b(\b+1)\bigr)^{3/2}\bigl[CU_{(3)}^{-1}C_{(3)}\bigr]_n
\left(\bar{N}^3_{|n|}\right)^2\nonumber\\
&=\m g_2\l'\sqrt{-\b(\b+1)}\frac{\sin^2n\pi\b}{2\pi^2}\left[1-\frac{1}{4}(3-16a_R)
\l'n^2+{\mc O}(\l^{\prime\,2})\right]\,,\\
\label{fer2}
g_{\text{s}}\la n_{5,6}|H_3|m_{7,8}\ra|v\ra & =
-\frac{1}{2}\m g_2\a_1^2\a_2^2(\b+1)(1-4\m\a K)^{-2}
\bigl[CU_{(1)}^{-1}C_{(1)}\bigr]_m\bigl[CU_{(3)}^{-1}C_{(3)}\bigr]_n
\left(\bar{N}^1_{|m|}\bar{N}^3_{|n|}\right)^2\nonumber\\
&=\m g_2\l'(\b+1)\frac{\sin^2n\pi\b}{2\pi^2}+{\mc 
O}(\l^{\prime\,2})\,.
\end{align}
Up to a sign, the first amplitude in equation \eqref{fer1} is identical to its bosonic analogue \eqref{00n1} for all 
$\m$, whereas the second one, equation \eqref{fer2}, is the same at leading order. These results reflect that the 
vertex constructed in section~\ref{sec4} satisfies the superalgebra.
One can check
that equations~\eqref{fer1} and~\eqref{fer2} are unchanged if all spinor 
indices are in $\{1,2,3,4\}$ and (up to permutations) vanish otherwise.
 
\vskip1cm
\section*{Acknowledgement}
We are grateful for discussions with G.~Arutyunov, N.~Beisert, B.~de Wit, M.~B.~Green, S.~Minwalla, 
A.~Petkou, F.~Quevedo and M.~Taylor. In particular, we would like to thank J.~Plefka and 
S.~Theisen for many useful comments on the manuscript and for sharing their insights. 
Part of this work was performed during the workshop ``The quantum structure of 
spacetime and the geometric nature of fundamental interactions'' in Leuven. We would like to thank the organizers 
for the stimulating atmosphere they created. 
This work was supported by GIF, the German-Israeli foundation
for Scientific Research, FOM, the Dutch Foundation for Fundamental Research on Matter and 
the European Community's Human Potential Programme under contract HPRN-CT-2000-00131 
in which A.~P.~is associated to the University of Bonn.  
\appendix
\section{Some computational details}\label{appB}
\setcounter{section}{1}
\renewcommand{\thesection}{\Alph{section}}
\setcounter{equation}{0}
Here we give the details leading to equations \eqref{qk}, \eqref{qy}, \eqref{qv} and
\eqref{tqv}. The explicit expression for the infinite component vector $F_{(r)}$ appearing
in the bosonic constituent of the prefactors is
\begin{equation}\label{f+}
F_{(r)}=-\frac{1}{\sqrt{\a'}}\frac{\a}{1-4\m\a K}\frac{1}{\a_r}
U_{(r)}^{-1}C_{(r)}^{1/2}C\bar{N}^r\,.
\end{equation}
The fermionic quantity $G_{(r)}$ can be expressed in terms of $F_{(r)}$ as \cite{ap}
\begin{equation}\label{gf}
G_{(r)}=\bigl(1-2\m\a K(1-\Pi)\bigr)\sqrt{|\a_r|}P_{(r)}^{-1}U_{(r)}C^{-1/2}F_{(r)}\,.
\end{equation}
The following identities prove very useful ($\a_3\Theta\equiv\vt_{0(1)}-\vt_{0(2)}$)
\begin{align}
\label{r}
\mathbb{R}|V\ra & =i\sqrt{\a'}\left[2K\sqrt{\a'}
\left(\mathbb{P}-i\frac{\m\a}{\a'}\mathbb{R}\right)+
\sum_{r,n>0}C_{n(r)}^{1/2}\bar{N}^r_na^{\dg}_{n(r)}\right]|V\ra\,,\\
\label{theta}
\Theta|V\ra & =-\sqrt{2}\sum_{r,n}Q^r_nb_{-n(r)}^{\dg}|V\ra\,.
\end{align}
Using the mode expansions of $Q^-_{(r)}$, $\bar{Q}^-_{(r)}$,
${\mc K}_0+{\mc K}_+$, ${\mc K}_-$ and ${\mc Y}$ given in equations
\eqref{q-mode}, \eqref{k+mode}, \eqref{k-mode} and \eqref{ymode} one gets
\begin{align}
\sum_{r=1}^3\{Q^-_{(r)},{\mc Y}\} & =
-\g\sum_{r=1}^3\frac{1}{\sqrt{|\a_r|}}\sum_{n=1}^{\infty}\bigl[P_{(r)}C^{1/2}G_{(r)}\bigr]_na_{-n(r)}^{\dg}\,,\\
\sum_{r=1}^3\{\bar{Q}^-_{(r)},{\mc Y}\} & = \mathbb{P}\g-i\frac{\m\a}{\a'}\mathbb{R}\g\Pi
+\g\sum_{r=1}^3\frac{1}{\sqrt{|\a_r|}}\sum_{n=1}^{\infty}\bigl[P_{(r)}^{-1}C^{1/2}G_{(r)}\bigr]_na_{n(r)}^{\dg}\,,
\end{align}
\begin{align}
\sum_{r=1}^3[Q^-_{(r)},{\mc K}_0+{\mc K}_+] & = \m\g(1+\Pi)\sqrt{\frac{2}{\a'}}\L+
\g\sum_{r=1}^3\frac{e(\a_r)}{\sqrt{|\a_r|}}\sum_{n=1}^{\infty}\bigl[P_{(r)}^{-1}C^{1/2}F_{(r)}\bigr]_nb_{n(r)}^{\dg}\,,\\
\sum_{r=1}^3[Q^-_{(r)},{\mc K}_-] & = i\g\sum_{r=1}^3\frac{e(\a_r)}{\sqrt{|\a_r|}}
\sum_{n=1}^{\infty}\bigl[P_{(r)}^{-1}C^{1/2}U_{(r)}F_{(r)}\bigr]_nb_{-n(r)}^{\dg}\,,\\
\sum_{r=1}^3[\bar{Q}^-_{(r)},{\mc K}_0+{\mc K}_+] & = -\frac{\m\a}{\sqrt{2\a'}}\g(1-\Pi)\Theta+
\g\sum_{r=1}^3\frac{e(\a_r)}{\sqrt{|\a_r|}}\sum_{n=1}^{\infty}\bigl[P_{(r)}C^{1/2}F_{(r)}\bigr]_nb_{-n(r)}^{\dg}\,,\\
\sum_{r=1}^3[\bar{Q}^-_{(r)},{\mc K}_-] & =-i\g\sum_{r=1}^3\frac{e(\a_r)}{\sqrt{|\a_r|}}
\sum_{n=1}^{\infty}\bigl[P_{(r)}C^{1/2}U_{(r)}F_{(r)}\bigr]_nb_{n(r)}^{\dg}\,.
\end{align}
Using equations \eqref{gf}, \eqref{r} and \eqref{theta} we find equations
equations \eqref{qk} and \eqref{qy}. The action of the supercharges on $|V\ra$ given in equations
\eqref{qv} and \eqref{tqv} can be proven similarly. One needs
\begin{equation}
\begin{split}
\bar{N}^{rs}_{nm}+e(\a_s)\left(\frac{m}{n}\left|\frac{\a_r}{\a_s}\right|\right)^{3/2}
P_{n(r)}P_{m(s)}Q^{rs}_{nm} & = -\frac{\a}{\a_s}(1-4\m\a K)^{-1}
\bigl[C_{(r)}^{1/2}\bar{N}^r\bigr]_n\bigl[U_{(s)}^{-1}C_{(s)}^{1/2}C\bar{N}^s\bigr]_m\,,\\
\bar{N}^{rs}_{-n,-m}+e(\a_r)\left(\frac{m}{n}\left|\frac{\a_r}{\a_s}\right|\right)^{1/2}P_{n(r)}P_{m(s)}Q^{rs}_{nm} & 
= 0\,,
\end{split}
\end{equation}
\begin{equation}
\begin{split}
\bar{N}^{rs}_{nm}-e(\a_r)\left(\frac{m}{n}\left|\frac{\a_r}{\a_s}\right|\right)^{1/2}P_{n(r)}^{-1}P_{m(s)}^{-1}Q^{rs}_{nm} & =
-\m\a(1-4\m\a K)^{-1}(1-\Pi)\bigl[C_{(r)}^{1/2}\bar{N}^r\bigr]_n\bigl[C_{(s)}^{1/2}\bar{N}^s\bigr]_m\,,\\
\bar{N}^{rs}_{-n,-m}-e(\a_s)\left(\frac{m}{n}\left|\frac{\a_r}{\a_s}\right|\right)^{3/2}P_{n(r)}^{-1}P_{m(s)}^{-1}Q^{rs}_{nm} & =
\frac{\a}{\a_s}(1-4\m\a K)^{-1}
\bigl[P^{-2}_{(r)}C_{(r)}^{1/2}\bar{N}^r\bigr]_n\bigl[C_{(s)}^{1/2}C\bar{N}^s\bigr]_m
\end{split}
\end{equation}
which follow from \eqref{nnpp} and \eqref{qmn}.
\section{Proof of equations \eqref{ppwave1} and \eqref{ppwave2}}\label{appD}
\setcounter{section}{2}
\renewcommand{\thesection}{\Alph{section}}
\setcounter{equation}{0}
In this appendix we prove that
\begin{align}
\label{1}
\g^I_{a(\dot{a}}\bigl[\Pi\bar{D}\bigr]^as^I_{\dot{b})} &=0\,,\\
\label{2}
\g^I_{a(\dot{a}}\bigl[\Pi D\bigr]^a\wt{s}^I_{\dot{b})} & =0\,,\\
\label{3} \Bigl(\g^I_{a\dot{a}}\bar{D}_b\wt{s}^I_{\dot{b}}+
\g^I_{a\dot{b}}D_bs^I_{\dot{a}}\Bigr)(1-\Pi)^{ab} & =0\,.
\end{align}
Equations~\eqref{1} and~\eqref{2} are equivalent to
\begin{align}
\label{1e}   
\Bigl(\g^I_{a(\da}Y_bs^I_{1\,\db)}
+\frac{\a}{\a'}\g^I_{a(\da}\frac{\p}{\p Y^b}s^I_{2\,\db)}\Bigr)
\Pi^{ab} & =0\,,\\
\label{2e}
\Bigl(\g^I_{a(\da}Y_bs^I_{2\,\db)}
-\frac{\a}{\a'}\g^I_{a(\da}\frac{\p}{\p Y^b}s^I_{1\,\db)}\Bigr)
\Pi^{ab} & =0\,,
\end{align}
The first equation has terms of order ${\mc O}(Y^2)$ and ${\mc O}(Y^6)$,
whereas the second one
has terms of order ${\mc O}(Y^0)$, ${\mc O}(Y^4)$ and ${\mc O}(Y^8)$.
There are two contributions to the order ${\mc O}(Y^2)$ in
equation~\eqref{1e}, both vanish separately.
The first one is
\begin{equation}
\g^I_{a(\da}Y_bs_{1\,\db)}^I\Pi^{ab}=
2\g^I_{a(\da}\g^I_{c\db)}Y^bY^c\Pi^{ab}=
-2\d_{\da\db}\Pi_{ab}Y^aY^b=0\,,
\end{equation}
whereas the second one is
\begin{align}
&\frac{\a}{\a'}\g^I_{a(\da}\frac{\p}{\p
Y^b}s_{2\,\db)}^I\Pi^{ab}=
-\g^I_{a(\da}u^I_{bcd\db)}Y^cY^d\Pi^{ab}=\nonumber\\
&\frac{1}{16}\left(\g^{IJ}\g^{KL}\right)_{(\da\db)}
\g^{IJ}_{a[b}\g^{KL}_{cd]}\Pi^{ab}Y^cY^d=
\frac{1}{24}\left(\g^{IJ}\g^{KL}\right)_{(\da\db)}
\left(\g^{IJ}\Pi\g^{KL}\right)_{cd}Y^cY^d=0\,.
\end{align}
Here we have used equations \eqref{gu} and \eqref{klpij}. From the
Fourier identities \cite{gsb}
\begin{equation}\label{fourier}
\begin{split}
s_{1\,\da}(\phi) & = \left(\frac{\a}{\a'}\right)^4\int
d^8Y\,s_{2\,\da}^I(Y)
e^{\frac{\a'}{\a}\phi Y}\,,\\
s_{2\,\da}(\phi) & = \left(\frac{\a}{\a'}\right)^4\int
d^8Y\,s_{1\,\da}^I(Y) e^{\frac{\a'}{\a}\phi Y}\,,
\end{split}
\end{equation}
it follows that the terms of order ${\mc O}(Y^6)$ vanish as well.
This proves equation \eqref{1}.
The ${\mc O}(Y^0)$ term in equation~\eqref{2e} is
\begin{equation}
\g^I_{a(\da}\g^I_{b\db)}\Pi^{ab}=\d_{\da\db}\tr(\Pi)=0\,,
\end{equation}
and the order ${\mc O}(Y^8)$ term vanishes by \eqref{fourier}.
The terms of order ${\mc O}(Y^4)$ in equation \eqref{2e} are
\begin{align}
&{\Pi^a}_b\g^I_{a(\da}u^I_{cde\db)}
\left(Y^bY^cY^dY^e+\frac{1}{24}{\e^{cdeb}}_{ghij}Y^gY^hY^iY^j\right)\nonumber 
\\
&=-\frac{1}{16}{\Pi^a}_b\left(\g^{IJ}\g^{KL}\right)_{(\dot{a}\dot{b})}
\g^{IJ}_{a[c}\g^{KL}_{de]}\left(Y^bY^cY^dY^e+\frac{1}{24}{\e^{cdeb}}_{ghij}
Y^gY^hY^iY^j\right)
\nonumber\\
&=-\frac{1}{16}{\Pi^a}_b\left(\g^{IJKL}_{\da\db}-2\d_{\da\db}\d^{IK}\d^{JL}\right)
\g^{IJ}_{a[c}\g^{KL}_{de]}\left(Y^bY^cY^dY^e+\frac{1}{24}{\e^{cdeb}}_{ghij}
Y^gY^hY^iY^j\right)
\nonumber\\
&=-\frac{1}{16}{\Pi^a}_b\g^{IJKL}_{\da\db}t^{IJKL}_{acde}
\left(Y^bY^cY^dY^e+\frac{1}{24}{\e^{cdeb}}_{ghij}Y^gY^hY^iY^j\right)=0\,.
\end{align}
In the last step we used that $\Pi$ is symmetric and traceless and
\begin{equation}
t^{IJKL}_{abcd}=-\frac{1}{24}{\e_{abcd}}^{efgh}t^{IJKL}_{efgh}\,.
\end{equation}
This proves equation \eqref{2}. Finally, equation \eqref{3} is
equivalent to
\begin{align}
\label{sym3} &
\Bigl(\g^I_{a(\da}Y_bs_{1\,\db)}^I-\frac{\a}{\a'}\g^I_{a(\da}\frac{\p}{\p
Y^b}s_{2\,\db)}^I
\Bigr)(1-\Pi)^{ab}=0\,,\\
\label{asym3} &
\Bigl(\g^I_{a[\da}Y_bs_{2\,\db]}^I+\frac{\a}{\a'}\g^I_{a[\da}\frac{\p}{\p
Y^b}s_{1\,\db]}^I \Bigr)(1-\Pi)^{ab}=0\,.
\end{align}
The first equation is symmetric in $\da$, $\db$ and contains terms
of order ${\mc O}(Y^2)$ and ${\mc O}(Y^6)$. These vanish for the
same reason as those in equation \eqref{1}. The second equation is
antisymmetric in $\da$, $\db$ and contains terms of order ${\mc O}(Y^0)$, ${\mc O}(Y^4)$
and ${\mc O}(Y^8)$.
The ${\mc O}(Y^0)$ contribution to equation \eqref{asym3} is
\begin{equation}
\g^I_{a[\da}\g^I_{b\db]}(1-\Pi)^{ab}=\frac{1}{4}\g^{IJ}_{\da\db}\g^I_{ab}(1-\Pi)^{ab}=0\,.
\end{equation}
From equation \eqref{fourier} it follows that the term of order
${\mc O}(Y^8)$ vanishes as well. Finally, there are two
contributions to the terms of order ${\mc O}(Y^4)$, both of them
vanish separately. The first one is
\begin{align}
&\frac{\a}{\a'}\g^I_{a[\da}Y_bs_{2\,\db]}^I(1-\Pi)^{ab}=
-\frac{1}{3}\g^I_{a[\da}u^I_{cde\db]}{(1-\Pi)^a}_bY^bY^cY^dY^e=\nonumber\\
&\frac{1}{12}\left(\g^{IJ}_{\da\db}\d_{a[c}\g^{IJ}_{de]}
+\frac{1}{4}\left(\g^{IJ}\g^{KL}\right)_{[\da\db]}\g^{IJ}_{a[c}\g^{KL}_{de]}\right)
{(1-\Pi)^a}_bY^bY^cY^dY^e=\nonumber\\
&\frac{1}{12}\g^{IJ}_{\da\db}\g^{IK}_{a[c}\g^{KJ}_{de]}{(1-\Pi)^a}_bY^bY^cY^dY^e=
\frac{1}{6}\g^{IJ}_{\da\db}\g^{IJ}_{bc}(1-\Pi)_{de}Y^bY^cY^dY^e=0\,.
\end{align}
In the last step we have used equation \eqref{ij}. The second
contribution of order ${\mc O}(Y^4)$ then vanishes by equation
\eqref{fourier}. This concludes the proof of equation
\eqref{asym3}.

\noindent Apart from symmetry and tracelessness of $\Pi$ we have
used the following identities
\begin{align}
\g^{IJ}_{ab}&=-\g^{IJ}_{ba}\,,\\
\g^I_{a\da}\g^I_{b\db}&=\d_{ab}\d_{\da\db}+\frac{1}{4}\g^{IJ}_{ab}\g^{IJ}_{\da\db}\,,\\
\left(\g^{IJ}\g^{KL}\right)_{ab}&=\g^{IJKL}_{ab}+\d^{IL}\g^{JK}_{ab}+\d^{JK}\g^{IL}_{ab}
\nonumber \\
&-\d^{IK}\g^{JL}_{ab}-\d^{JL}\g^{IK}_{ab}+\left(\d^{JK}\d^{IL}-\d^{JL}\d^{IK}\right)\d_{ab}
\,,\\
\label{gu}
\g^I_{a\da}u^I_{bcd\db}&=-\frac{1}{4}\g^{IJ}_{\da\db}\d_{a[b}\g^{IJ}_{cd]}
-\frac{1}{16}\left(\g^{IJ}\g^{KL}\right)_{\da\db}\g^{IJ}_{a[b}\g^{KL}_{cd]}\,,\\
\label{ij}
\g^{IK}_{a[b}\g^{JK}_{cd]} & =t^{IJ}_{abcd}-2\d_{a[b}\g^{IJ}_{cd]}\,,\\
\label{ijij}
\g^{IJ}_{ab}\g^{IJ}_{cd} & = 8\bigl(\d_{ac}\d_{bd}-\d_{ad}\d_{bc}\bigr)\,,\\
\label{klpij}
\gamma^{IJKL}_{\da\db}\left(\g^{KL}\Pi\g^{IJ}\right)_{[ab]}&=0 \,.
\end{align}
\section{$\{Q,\wt{Q}\}$ at order ${\mc O}(g_{\text{s}})$}\label{appC}
\setcounter{section}{3}
\renewcommand{\thesection}{\Alph{section}}
\setcounter{equation}{0}
Here we demonstrate that equation \eqref{dyn3} leads to the constraints 
\eqref{qq1}--\eqref{ppwave2} given in section~\ref{sec4}.
To this end, we adopt a trick introduced in \cite{gsb}. Namely, we associate
the world-sheet coordinate dependence with the oscillators as
\begin{align}
a_{n(r)} &\to e^{-i\o_{n(r)}\t/\a_r}
\left(a_{n(r)}\cos\frac{n\s_r}{\a_r}-a_{-n(r)}\sin\frac{n\s_r}{\a_r}\right)\,,\\
a_{-n(r)} &\to e^{-i\o_{n(r)}\t/\a_r}
\left(a_{-n(r)}\cos\frac{n\s_r}{\a_r}+a_{n(r)}\sin\frac{n\s_r}{\a_r}\right)\,,\\
b_{n(r)} &\to e^{-i\o_{n(r)}\t/\a_r}
\left(b_{n(r)}\cos\frac{n\s_r}{\a_r}-b_{-n(r)}\sin\frac{n\s_r}{\a_r}\right)\,,\\
b_{-n(r)} &\to e^{-i\o_{n(r)}\t/\a_r}
\left(b_{-n(r)}\cos\frac{n\s_r}{\a_r}+b_{n(r)}\sin\frac{n\s_r}{\a_r}\right)\,.
\end{align}
Then we integrate the constraint equation \eqref{dyn3}
over the $\s_r$. In dealing with the resulting expressions we can
integrate by parts since the integrand is periodic.
In addition to the identities in equations \eqref{qy},\footnote{In fact, here
we need the analogue of equation \eqref{qy} with
${\mc K}\leftrightarrow\wt{\mc K}$.}
\eqref{qv} and \eqref{tqv} we have to calculate the
commutator of $\sum_rQ_{(r)}$ with ${\mc K}$ and and its tilded
counterpart. We get
\begin{equation}
\begin{split}
\sqrt{2}\eta &\sum_{r=1}^3[Q_{(r)},{\mc K}]\,|V\ra=-2i(1-4\m\a 
K)^{-1/2}\g
\left[\dot{Y}+Y'+\frac{i}{2}\m(1-\Pi)\left(Y-2Y_0\right)\right]|V\ra\,,\\
\sqrt{2}\bar{\eta}&\sum_{r=1}^3[\wt{Q}_{(r)},\wt{\mc 
K}]\,|V\ra=-2i
(1-4\m\a K)^{-1/2}\g\left[\dot{Y}-Y'+\frac{i}{2}\m(1-\Pi)\left(Y-2Y_0\right)\right]|V\ra\,.
\end{split}
\end{equation}
Here $Y_0$ is the zero-mode part of $Y$, we suppressed the $\t$, $\s_r$ dependence and
\begin{equation}
\dot{Y}\equiv\p_{\t}Y\,,\qquad Y'\equiv\sum_{r=1}^3\p_{\s_r}Y\,.
\end{equation}
The fact that the above equations have a term which only depends
on the zero mode $Y_0$ is important. Combining the various
contributions to equation \eqref{dyn3} yields
\begin{align}
\int & \prod_{r=1}^3d\s_r\,\Biggl( \sqrt{2}(1-4\m\a K)^{-1}\left[
\left(\g^I_{a\dot{a}}\eta\tilde{s}^I_{\dot{b}}-\g^{I}_{a\dot{b}}\bar{\eta}s^I_{\dot{a}}\right)\dot{Y}^a
+\left(\g^I_{a\dot{a}}\eta\tilde{s}^I_{\dot{b}}+\g^{I}_{a\dot{b}}\bar{\eta}s^I_{\dot{a}}\right){Y'}^a\right]
\nonumber\\
&-\frac{\m}{\sqrt{2}}(1-4\m\a K)^{-1}\Bigl[
\left(\g^I_{a\dot{a}}\bar{D}_b\tilde{s}^I_{\dot{b}}+\g^I_{a\dot{b}}D_bs^I_{\dot{a}}\right)(1-\Pi)^{ab}
+2i\left(\g^I_{a\dot{a}}\eta\tilde{s}^I_{\dot{b}}-\g^I_{a\dot{b}}\bar{\eta}s^I_{\dot{a}}\right){(1-\Pi)^a}_bY_0^b
\Bigr]\nonumber\\
&\label{qtq1}+\frac{i}{\sqrt{2}}\frac{\a'}{\a}\left[ {\mc K}^I{\mc
K}^J\g^J_{a\dot{a}}D^a\tilde{s}^I_{\dot{b}}-\wt{\mc K}^I\wt{\mc
K}^J\g^J_{a\dot{b}}\bar{D}^as^I_{\dot{b}}
\right]\Biggr)|V\ra=0\,.
\end{align}
Next we remove the $\s_r$ derivatives from $Y$ by partial
integration. This requires the identity~\cite{gsb}\footnote{There are some minor typos in appendices 
D and E of~\cite{gsb} which we correct for the interested reader to facilitate the comparison with this paper. 
The right-hand-side of equation (D.25) in~\cite{gsb} should be multiplied by $\frac{1}{2\a}$; on the 
r.h.s. of equations (D.26) and (D.27) it should be $\frac{1}{2^{3/2}}$ instead of $\frac{1}{\sqrt{2}}$; the 
r.h.s. of equation (E.8) should be multiplied by $\frac{1}{\a}$. This corrections modify equation (E.9) of~\cite{gsb}
slightly, which is now in agreement with equations~\eqref{qq1}--\eqref{qq3} of this paper. 
These typos do not affect the proof presented in appendix E of~\cite{gsb}. }
\begin{equation}
\Bigl(\g^I_{a\dot{a}}\eta\tilde{s}^I_{\dot{b}}+\g^I_{a\dot{b}}\bar{\eta}s^I_{\dot{a}}\Bigr){Y'}^a
=-\frac{2^{3/2}\a}{\a'}m'_{\dot{a}\dot{b}}
\end{equation}
and
\begin{equation}
\sum_{r=1}^3\p_{\s_r}|V\ra=-\frac{i}{4}\frac{\a'}{\a}\Bigl(
(1-4\m\a K)\bigl({\mc K}^2-\wt{\mc
K}^2\bigr)+4\bigl(Y\dot{Y}+i\m(1-\Pi)YY_0\bigr)\Bigr)|V\ra\,.
\end{equation}
Using the above identities equation \eqref{qtq1} becomes after
integrating by parts
\begin{equation}
\begin{split}
\int\prod_{r=1}^3d\s_r\,\Biggl(& (1-4\m\a K)^{-1}\Bigl[
\sqrt{2}\bigl(\g^I_{a\dot{a}}\eta\tilde{s}^I_{\dot{b}}-\g^{I}_{a\dot{b}}
\bar{\eta}s^I_{\dot{a}}\bigr)
-4im_{\dot{a}\dot{b}}Y_a\Bigr]\Bigl(\dot{Y}^a-\dot{Y}^a_0\Bigr)\\
&-\frac{\m}{\sqrt{2}}(1-4\m\a K)^{-1}
\Bigl(\g^I_{a\dot{a}}\bar{D}_b\tilde{s}^I_{\dot{b}}+\g^I_{a\dot{b}}D_b
s^I_{\dot{a}}\Bigr)(1-\Pi)^{ab}
\\
&-i{\mc K}^I{\mc 
K}^J\Bigl[\d^{IJ}m_{\dot{a}\dot{b}}-\frac{\a'}{\sqrt{2}\a}
\g^J_{a\dot{a}}D^a\tilde{s}^I_{\dot{b}}\Bigr]
+i\wt{\mc K}^I\wt{\mc 
K}^J\Bigl[\d^{IJ}m_{\dot{a}\dot{b}}-\frac{\a'}{\sqrt{2}   
\a}\g^J_{a\dot{b}}
\bar{D}^as^I_{\dot{a}}\Bigr]\Biggr)|V\ra=0\,.
\end{split}   
\end{equation}
\section{The Functional Approach to the Prefactor}\label{appf}
\setcounter{section}{4}
\renewcommand{\thesection}{\Alph{section}}
\setcounter{equation}{0}
The bosonic constituents of the prefactor can be obtained via the operators~\cite{gs,gsb}
\begin{equation}
\begin{split}
\p X(\s) & =
4\pi\frac{\sqrt{-\a}}{\a'}(\pi\a_1-\s)^{1/2}\bigl(\grave{x}_1(\s)+\grave{x}_1(-\s)\bigr)\,,\\
\label{p}
P(\s) &=-2\pi\sqrt{-\a}(\pi\a_1-\s)^{1/2}\bigl(p_1(\s)+p_1(-\s)\bigr)\,.
\end{split}
\end{equation}
Acting on the exponential part of the vertex these satisfy
\begin{equation}\label{pp}
\begin{split}
\lim_{\s\to\pi\a_1}K(\s)|V\ra & \equiv
\lim_{\s\to\pi\a_1}\left(P(\s)+\frac{1}{4\pi}\p X(\s)\right)|V\ra
=f(\m){\mc K}|V\ra\,,\\
\lim_{\s\to\pi\a_1}\wt{K}(\s)|V\ra & \equiv
\lim_{\s\to\pi\a_1}\left(P(\s)-\frac{1}{4\pi}\p X(\s)\right)|V\ra
=f(\m)\wt{\mc K}|V\ra\,.
\end{split}
\end{equation}
Here we defined
\begin{equation}\label{fmu}
f(\m)\equiv-2\frac{\sqrt{-\a}}{\a_1}\lim_{e\to0}\e^{1/2}\sum_{n=1}^{\infty}(-1)^nn\cos(n\e/\a_1)\bar{N}^1_n
\end{equation}
which is equal to one for $\m=0$ \cite{gs}. For the fermionic constituent of the prefactor
one considers~\cite{gs,gsb}
\begin{equation}
Y(\s)=-2\pi\frac{\sqrt{-2\a}}{\sqrt{\a'}}(\pi\a_1-\s)^{1/2}\bigl(\l_1(\s)+\l_1(-\s)\bigr)
\end{equation}
which satisfies \cite{ap}
\begin{equation}
\lim_{\s\to\pi\a_1}Y(\s)|V\ra=f(\m)(1-4\m\a K)^{-1}(1-2\m\a K(1+\Pi)){\mc Y}|V\ra\,.
\end{equation}
Comparing this identity with equation \eqref{y}, we see that
\begin{equation}
\lim_{\s\to\pi\a_1}Y(\s)|V\ra=Y|V\ra\,,\qquad\Longleftrightarrow\qquad
f(\m)=(1-4\m\a K)^{1/2}\,.
\end{equation}
We conjecture that this is indeed the case. Unfortunately, it seems that there is no way
of proving this conjecture, at least for large-$\m$, without knowing $\bar{N}^r$
for all $\m$. This is because in equation~\eqref{fmu}, the large-$\m$ expansion does not commute with the infinite sum 
over $n$. 

\noindent One can write the functional equivalent of equations
\eqref{h}, \eqref{q} and \eqref{tq} as 
\begin{align}
H_3 & = \lim_{\s\to\pi\a_1}\int d\m_3
\Bigl(\wt{K}^I(\s)K^J(\s)-\frac{\m\a}{\a'}\d^{IJ}\Bigr)v_{IJ}(Y(\s))\Phi(1)\Phi(2)\Phi(3)\,,\\
Q_{3\,\da} & = \lim_{\s\to\pi\a_1}\int 
d\m_3\wt{K}^I(\s)s^I_{\da}(Y(\s))\Phi(1)\Phi(2)\Phi(3)\,,\\
\wt{Q}_{3\,\da} & = \lim_{\s\to\pi\a_1}\int 
d\m_3K^I(\s)\wt{s}^I_{\da}(Y(\s))\Phi(1)\Phi(2)\Phi(3)\,.
\end{align}
Here
\begin{equation}
d\m_3\equiv \left(\prod_{r=1}^3d\a_rD^8\l_r(\s)D^8p_r(\s)\right)
\d\bigl(\sum_s\a_s\bigr)\D^8\bigl[\sum_s\l_s(\s)\bigr]\D^8\bigl[\sum_sp_s(\s)\bigr]
\end{equation}
is the functional expression leading to the exponential part of the vertex \cite{gs,gsb}.

\noindent Finally, let us mention the following subtlety. Using equation \eqref{qfield} it is
easy to see that for example 
\begin{equation}
\sqrt{2}\bar{\eta}\sum_{r=1}^3[\wt{Q}_{(r)},\lim_{\s\to\pi\a_1}K(\s)]\,|V\ra=
\m\g\Pi Y|V\ra\,.
\end{equation}
However, this is not equal to the commutator of $\sum_r\wt{Q}_{(r)}$ with $f(\m){\mc K}$. To see this,
rewrite
\begin{align}
&\sqrt{2}\bar{\eta}\sum_{r=1}^3[\wt{Q}_{(r)},f(\m){\mc K}]|V\ra=
\sqrt{2}\bar{\eta}\sum_{r=1}^3[\wt{Q}_{(r)},\lim_{\s\to\pi\a_1}K(\s)]\,|V\ra
\nonumber\\
\label{qkfunc}
&-\frac{\a'}{\a}\g Y(1-4\m\a K)f^{-1}(\m)[\lim_{\s\to\pi\a_1}K(\s),\wt{K}]|V\ra=
\m\g\Pi Y|V\ra+\m\g Y|V\ra=\m\g(1+\Pi)Y|V\ra\,.
\end{align}
Here we used equation \eqref{tqv} and 
\begin{equation}
[\lim_{\s\to\pi\a_1}K(\s),\wt{K}]|V\ra=
-\frac{\m\a}{\a'}(1-4\m\a K)^{-1}f(\m)|V\ra\,.
\end{equation}
Equation \eqref{qkfunc} is equivalent to equation
\eqref{qk} of section \ref{sec4}. It is this appearance of the matrix $1+\Pi$ as opposed to
just $\Pi$, that is responsible for the term proportional to $\m\d^{IJ}$ in the cubic
interaction vertex.
\section{Large-$\m$ Expansions}\label{appE}
\setcounter{section}{5}
\renewcommand{\thesection}{\Alph{section}}
\setcounter{equation}{0}
In this appendix we collect some results on the large-$\m$ expansion.
We need the explicit expressions for the matrices $A^{(r)}$ and the vector $B$.
For $m,n>0$ they are
\begin{equation}
\begin{split}
A^{(1)}_{mn} & = (-1)^n\frac{2\sqrt{mn}\b}{\pi}\frac{\sin m\pi\b}{m^2\b^2-n^2}\,,\\
A^{(2)}_{mn} & = \frac{2\sqrt{mn}(\b+1)}{\pi}\frac{\sin m\pi\b}{m^2(\b+1)^2-n^2}\,,\\
A^{(3)}_{mn} & = \d_{mn}\,,\\
B_m & =-\frac{2}{\pi}\frac{\a_3}{\a_1\a_2}m^{-3/2}\sin m\pi\b\,.
\end{split}
\end{equation}
Writing
\begin{equation}
\G^{-1}=\frac{1}{2}CC_{(3)}^{-1}+R
\end{equation}
it was shown in \cite{sch} that
\begin{equation}
R = a_R\frac{\pi}{(\m\a_3)^4}\left(\frac{\a_1\a_2}{\a_3}\right)^2C^3BB^TC^3+\cdots
\end{equation}
with $a_R$ an undetermined number. The next term in the expansion would be of order $\m^{-6}$.
It follows that
\begin{equation}
\bar{N}^3 = \frac{1}{\m\a_3}\left[-\frac{1}{2}C^{1/2}+
\frac{1}{(\m\a_3)^2}\left(\frac{1}{4}-a_R\right)C^{5/2}+\ldots\right]B\,.
\end{equation}
Furthermore the large-$\m$ expansion of $1-4\m\a K$ is~\cite{sch}
\begin{equation}
1-4\m\a K = \frac{a_K}{\pi\m\a_3}\frac{1}{\b(\b+1)}+\cdots\,.
\end{equation}
The next term in the expansion would be of order $\m^{-3}$. Here, $a_K$ is undetermined as
well, however it was shown in \cite{sch} that
\begin{equation}
a_Ra_K=\frac{1}{64}\,.
\end{equation}
From the definition of $\bar{N}^{rs}_{mn}$ it is easy to see that
\begin{equation}
\bar{N}^{r3}_{mn} \sim -\frac{2}{\pi}(-1)^{(m+1)r}\left(-\frac{\a_r}{\a_3}
\right)^{3/2}
\frac{n\sin 
n\pi\b}{n^2\left(\displaystyle{\frac{\a_r}{\a_3}}\right)^2-m^2}
\end{equation}
at leading order. Here and below $r$, $s\in\{1,2\}$.
Using the factorization theorem equation \eqref{nnpp}, the above formulae are 
sufficient to
determine the leading large-$\m$ behavior of all the other Neumann matrices 
as well. We find
\begin{equation}
\bar{N}^r_n \sim (-1)^{(n+1)r}\frac{a_K}{\pi}\frac{1}{\m^2\a_1\a_2\a_r}
\end{equation}
and therefore
\begin{equation}
\bar{N}^{rs}_{mn} \sim 
(-1)^{(m+1)r+(n+1)s}\frac{2a_K}{\pi\m\sqrt{\a_r\a_s}}\,,
\qquad
\bar{N}^{33}_{mn} \sim \frac{2a_K}{\pi(\m\a_3)^3}\bigl(m\sin m\pi\b\bigr)
\bigl(n\sin n\pi\b\bigr)\,.
\end{equation} 

\end{document}